%Paper: 9202002
%From: chung@beauty.tn.cornell.edu (Stephen Chung)
%Date: Sun, 2 Feb 92 18:35:46 EST

%%%%%%%%%%%%%%%%%%%%%%%%%%%%%%%%%%%%%%%%%%%%%%%%%%%%%%%%%%%%%%%%%%%%%
%%      Title:  Chiral Gauged WZW Theories and Coset Models     %%%%%
%%                in  Conformal Field Theory                    %%%%%
%%      Author: Stephen-wei Chung and S.-H. Henry Tye           %%%%%
%%              Newman Laboratory of Nuclear Studies            %%%%%
%%              Cornell University                              %%%%%
%%              Ithaca, NY 14853, USA                           %%%%%
%%      e-mail address: chung@LNSSUN1.CORNELL.EDU               %%%%%
%%                      chung@CRNLNUC.bitnet                    %%%%%
%%      Packages Used:  harvmac.tex                             %%%%%
%%                      pictex.tex (available in hepth)         %%%%%
%%%%%%%%%%%%%%%%%%%%%%%%%%%%%%%%%%%%%%%%%%%%%%%%%%%%%%%%%%%%%%%%%%%%%
\input harvmac.tex
\input pictex.tex
\overfullrule=0pt
\tolerance = 10000
\noblackbox

\font\eightrm=cmr8
\newfam\scfam \def\scfont{\fam\scfam\tenrm} \textfont\scfam=\eightrm
                                            \scriptfont\scfam=\fiverm

\font\twelvemi=cmmi12
%%%%%%%%%%%%%% Font defined in the Figure Caption inside the text.%%%%%
\font\fninerm=cmr9\font\fninei=cmmi9\font\fnineit=cmti9\font\fninesy=cmsy7
\font\fninebf=cmbx9\font\fninesl=cmsl9
\def\figufont{\def\rm{\fam0\fninerm}% switch back to 10-point type
\textfont0=\fninerm \scriptfont0=\sevenrm \scriptscriptfont0=\fiverm
\textfont1=\fninei  \scriptfont1=\seveni  \scriptscriptfont1=\fivei
\textfont2=\fninesy
\scriptfont2=\sevensy \scriptscriptfont2=\fivesy
\textfont\itfam=\fnineit \def\it{\fam\itfam\fnineit}
                                        \def\sl{\fam\slfam\fninesl}
\textfont\bffam=\fninebf \def\bf{\fam\bffam\fninebf} \rm}
%%%%%%%%%%%%%%%%%%%%%%%%%%%%%%%%%%%%%%%%%%%%%%%%%%%%%%%%%%%%%%%%

\def\Dirac#1{ #1 \hskip-6pt /\,}
\def\APsi{\bar \Psi}
\def\Apair{A^L,A^R}
\def\var{\delta_{v_L,v_R}}
\def\vraised{ {\raise1pt\hbox{\font\bigtenrm=cmr10 scaled\magstep1 $v$}} }
\def\vpair{\vraised_L,\vraised_R}
\def\anom{\omega^0_{2n+1}}
\def\df{\font\bigtenrm=cmbx12 scaled\magstep2 {\bf d}}
\def\smaDelta{{\scfont\Delta} }
\def\ie{{\it i.e.}}
\def\eg{{\it e.g.}}
\def\a{\alpha}
\def\b{\beta}
\def\c{\gamma}
\def\d{\delta}
\def\KM{ ${\rm Ka}{\check {\rm c}}{\rm -Moody}$ }
\def\o{ \omega}
\def\do{ \omega^{\dagger} }
\def\e{\epsilon}
\def\SU{SU(2)_k\otimes SU(2)_l/SU(2)_{k+l}}
%%%%%%%%%%%%%%%%%%%%%%%%%%%%%%%%%%  REFERENCE Abbreviation %%%%%%%%%%%%%
\def\IR{\relax{\rm I\kern-.18em R}}
\def\npb#1#2#3{{\it Nucl.\ Phys.} {\bf B#1} (19#2) #3}
\def\plb#1#2#3{{\it Phys.\ Lett.} {\bf #1B} (19#2) #3}

\def\prl#1#2#3{{\it Phys.\ Rev.\ Lett.} {\bf #1} (19#2) #3}
\def\mpl#1#2#3{{\it Mod.\ Phys.\ Lett.} {\bf A#1} (19#2) #3}
\def\ijmp#1#2#3{{\it Int.\ Jour.\ Mod.\ Phys.} {\bf A#1} (19#2) #3}
\def\prd#1#2#3{{\it Phys.\ Rev.} {\bf D#1} (19#2) #3}
\def\pr0#1#2#3{{\it Phys.\ Rev.} {\bf #1} (19#2) #3}
\def\aop#1#2#3{{\it Annals\ of\ Phys.} {\bf #1} (19#2) #3}
\def\cmp#1#2#3{{\it Commun.\  Math.\ Phys.} {\bf #1} (19#2) #3}
\def\actap#1#2#3{{\it Acta.\ Physica\ Polonica} {\bf B#1} (19#2) #3}
\def\jmp#1#2#3{ {\it J.\ Math.\ Phys.} {\bf #1} (19#2) #3}
\def\jetp#1#2#3{ {\it Sov.\ Phys.\ J.E.T.P.} {\bf #1} (19#2) #3}
%%%%%%%%%%%%%%%%%%%%%%%%%%%%%%%%%%%%%%%%%%%%%%%%%%%%%%%%%%%%%%%%%%%%%%%%%

\Title{\vbox{\baselineskip12pt\hbox{CLNS 91/1127}}}
{{\vbox {\centerline{Chiral Gauged WZW Theories and Coset Models }
\vskip10pt\centerline{in Conformal Field Theory}   }}}
\bigskip

\centerline{{\it Stephen-wei  Chung}\footnote*{
E-mail address: \hbox{\vtop{ \hbox{chung@CRNLNUC.bitnet\hfil}
               \hbox{chung@LNSSUN1.tn.cornell.edu} }}\hfil }
{\it  and S.-H. Henry Tye}                                   }

\medskip
\centerline{Newman Laboratory of Nuclear Studies}
\centerline{Cornell University}
\centerline{Ithaca, N.Y.  14853-5001, USA}

\vskip.6in

\noindent
The Wess-Zumino-Witten (WZW) theory has a global symmetry denoted by
  $G_L\otimes G_R$.  In the standard gauged WZW theory, vector gauge fields
(\ie\ with vector gauge couplings) are
in the adjoint representation of the subgroup $H \subset G$.
In this paper, we show that, in the conformal limit in two dimensions, there
is a gauged WZW theory where the gauge fields are chiral and belong
to the subgroups $H_L$ and $H_R$ where $H_L$ and $H_R$ can be different groups.
In the special case where $H_L=H_R$, the theory is equivalent to vector gauged
WZW theory.  For general groups $H_L$ and $H_R$, an examination of the
correlation functions (or more precisely, conformal blocks)
 shows  that the chiral gauged WZW theory is equivalent to
$(G/H)_L\otimes (G/H)_R$ coset models in conformal field theory.
The equivalence of the vector gauged WZW theory and the corresponding $G/H$
coset theory then follows.

%\vskip.2in
%\centerline{\bf Submitted to Nuclear Physics B.}
\vskip .4in

\Date{Jan, 1992}

%\draftmode
\lref\WZref{J.~Wess and B.~Zumino,~\plb{37}{71}{95}.}
\lref\EWone{E.~Witten,~\npb{223}{83}{422};~\npb{223}{83}{433}.}
\lref\Zumino{B.~Zumino, Y.-S. Wu and A. Zee,~\npb{239}{84}{477}\semi
             B.~Zumino, in ``Relativity, Groups and Topology II''
             Les Houches 1983, eds., B.S. DeWitt and R. Stora (North-Holland,
             Amsterdam, 1984).}
\lref\KT{H. Kawai and S.-H.H. Tye,~\plb{140}{84}{403}. }
\lref\Schone{H.J.~Schnitzer,~\npb{324}{89}{412}\semi
             D. Karabali, Q-Han Park, H.J.~Schnitzer and Z. Yang,
                                             ~\plb{216}{89}{307}\semi
             D. Karabali and H.J.~Schnitzer,~\npb{329}{90}{649}\semi
             D. Karabali, Talk given at NATO ARW 18th Int. Conf.
             on Differential Geometric Methods in Theoretical Physics:
             Physics and Geometry, Tahoe City, CA, July 2-8, 1989.}
\lref\GK{K. Gawedzki and A. Kupiainen,~\plb{215}{88}{119};
                                    ~\npb{320}{89}{625}\semi
         K. Gawedzki, in ``From Functional Integration, Geometry and Strings'',
         ed. by Z. Haba and J. Sobczyk (Birkhaeuser, 1989).}
\lref\EWfour{E. Witten, ``On Holomorphic Factorization of WZW and Coset
            Models'', IASSNS-HEP-91-25.}
\lref\EWtwo{E. Witten,~\cmp{92}{84}{455} }
\lref\KZref{V.G. Knizhnik and A.B. Zamolodchikov,~\npb{247}{84}{83}.}
\lref\BN{L.S. Brown and R.I. Nepomechie,~\prd{35}{87}{3239}\semi
         L.S. Brown, G.J. Goldberg, C.P. Rim and R.I. Nepomechie,
                                              ~\prd{36}{87}{551}.}
\lref\Po{A.M. Polyakov, \plb{103}{81}{207};~\plb{103}{81}{211}.}
\lref\PWREF{A.M. Polyakov and P.B. Wiegmann, \plb{131}{83}{121};
                                         ~\plb{141}{84}{223}.}
\lref\Alva{A. D'adda, A.C. Davis and P. Di Vecchia, \plb{121}{83}{335}\semi
           O. Alvarez, \npb{238}{84}{61}\semi
           R.I. Nepomechie, \aop{158}{84}{67}\semi
           P. Di Vecchia, B. Durhuus and J.L. Peterson, \plb{144}{84}{245}\semi
           K.D. Rothe, \npb{269}{86}{269}.}
\lref\Schtwo{A.N. Redlich and H.J. Schnitzer, \plb{167}{86}{315};
                                                 ~\plb{193}{87}{536}~(E).}
\lref\ZY{Z. Hlousek and K. Yamagishi, \plb{173}{86}{65}\semi
         T. Kawai, \plb{168}{86}{355}.}
\lref\Petersen{J.L. Petersen,~\actap{16}{85}{271}.}
\lref\SCM{T. Skyrme, {\it Proc. R. Soc. London}, {\bf A262} (1961) 237\semi
          S. Coleman,~\prd{11}{75}{2088};~\jmp{12}{71}{1735}\semi
          S. Mandelstam,~\prd{11}{75}{3026}. }
\lref\AG{L. Alvarez-Gaume and P. Ginsparg,
                           ~\npb{243}{84}{449};~\aop{161}{85}{423}.}
\lref\Jackiw{Extensive references can be found in
    ``Current Algebra and Anomalies'', ed. by R. Jackiw, S.B. Treiman,
      E. Witten, B. Zumino, World Scientific Pub., Singapore, 1986.}
\lref\EWthree{See, e.g., E. Witten,~\prd{44}{91}{314}.}
\lref\BARS{I. Bars,~\npb{334}{90}{125}\semi
I. Bars and  D. Nemeschansky, ~\npb{348}{91}{89}. }
\lref\GKOref{P. Goddard, A. Kent and D. Olive,~\cmp{103}{86}{105}.}
\lref\BPZ{A.A. Belavin, A.M. Polyakov, and A.B. Zamolodchikov,
                                         ~\npb{241}{84}{333}.}
\lref\Kacref{V.G. Ka{$\check {\rm c}$}, Infinite Dimensional Lie Algebras,
     3rd Ed. (Cambridge University Press, 1990).}

\lref\ddk{J. Distler and H. Kawai,~\npb{321}{89}{509}\semi
             F. David,~\mpl{3}{88}{1651}. }
\lref\Others{S. Das, S. Naik and S. Wadia,~\mpl{4}{89}{1033}\semi
             J. Polchinski,~\npb{324}{89}{123}\semi
             S. Das, A. Dhar and S. Wadia,~\mpl{5}{90}{799}\semi
             T. Banks and J. Lykken,~\npb{331}{90}{173}\semi
             A. Tseytlin,~\ijmp{5}{90}{1833}.  }
\lref\FMS{D. Friedan, E. Martinec and S. Shenker,~\npb{271}{86}{93}. }
\lref\WK{M. Wakimoto,~\cmp{104}{86}{605}. }
\lref\Fuji{K. Fujikawa,~\prl{42}{79}{1195};~\prd{21}{80}{2848}. }
\lref\AT{P.C. Argyres and S.-H.H. Tye,~\prl{67}{91}{3339}.}
\lref\Gepner{D. Gepner,~\npb{290}{87}{10}.}
\lref\ZFref{A.B. Zamolodchikov and V.A. Fateev,~\jetp{62}{85}{215}.}
\lref\ACT{P.C. Argyres, S.-w. Chung and S.-H.H. Tye, unpublished.}
\lref\KMQref{D. Kastor, E. Martinec and Z. Qiu,~\plb{200}{88}{434}\semi
   J. Bagger, D. Nemeschansky and S. Yankielowicz,~\prl{60}{88}{389}\semi
F. Ravanini,~\mpl{3}{88}{397}\semi
S.-w. Chung, E. Lyman and S.-H.H. Tye, Cornell preprint CLNS 91/1057,
{\it ``Fractional Supersymmetry and Minimal Coset Models
in Conformal Field Theory''} to appear in \ijmp{7}{92}{no.14}.}
\lref\AGT{P.C.~Argyres, J.~Grochocinski and S.-H.H.~Tye,~\npb{367}{91}{217} . }
\lref\keith{K.R.~Dienes and S.-H.H.~Tye, MCGILL-91-29,
{\it ``Model Building for Fractional Superstrings''}.}
%%%%%%%%%%%%%%%%%%%%%%%%%%%%%%%%%%%%%%%%%%%%%%%%%%%%%%%%%%%%%%%%%%%%%%%%%

%  TO TEST THE RIGHTHAND SIDE PAGE
%  \eject

\newsec{Introduction}

The topological term (\ie\ the Wess-Zumino (WZ) term\WZref ) in the
Wess-Zumino-Witten
(WZW) theory reflects the anomalies present in the theory of non-linear sigma
models.  Generically, the couplings of WZW actions to gauge fields  contain
anomalies.  For external
(background) gauge fields, this may be a desired feature, as in the case of
the two-photon decay of neutral pions.  If the gauge fields are to be treated
as real degrees of freedom (\ie\ to be integrated over in the functional
integration formulation), the consistency of the resulting gauge theory
requires such gauge couplings to be free from anomalies; that is, the theory
must be gauge-invariant.  In the absence of matter fields other than the
non-linear
sigma field, vector gauge couplings are automatically anomaly-free.
In fact, vector gauged WZW theories have been studied extensively in the
literature\refs{\EWone,\Schone,\GK}.

In this paper, we shall show that, in two dimensions (and in the conformal
limit which we are mainly interested in), there
exists another way to
introduce gauge couplings into the WZW theory that is also anomaly-free.
In this gauged WZW theory, the gauge fields are chiral, \ie\ each gauge field
has only one helicity.  In this type of gauged WZW theories, the holomorphic
(and anti-holomorphic) properties appear to be more transparent than in the
other gauged WZW theories.

This paper is organized as follows.  In Sec. 2, the coupling of the WZW
actions to gauge fields  and the
anomaly associated with it are briefly reviewed.  From the form of the
non-abelian gauge
anomaly, we can find special subsets of gauge couplings that are anomaly-free.
Besides the vector (or equivalently the axial) gauge couplings, we show that
there are two other types of gauge symmetries that are anomaly-free:

(1)  local \KM (affine KM) symmetry: in this case, the  local KM symmetry
arises from a gauge symmetry.  However, the gauge
fields happen to decouple from the sigma field, leaving behind the local
symmetry in the original (\ie\ ungauged) WZW model, which is precisely the
affine KM symmetry originally found  by Witten\refs{\EWtwo,\PWREF,\KZref}.

(2) chiral gauge symmetry: in this case the gauge fields in the group
$H_L\subset G_L$ have only the positive helicity along the left-moving
light-cone ($A^L_{\mu}= (0, A^L_{\bar z}$)) and the
gauge fields in the group $H_R\subset G_R$ have only the negative helicity
along the right-moving light-cone ($A^R_{\mu}=(A^R_{z}, 0$)).

This chiral gauged WZW theory is studied in Sec. 3.  The analysis is very
similar to the usual vector gauged WZW theory\refs{\Schone,\GK}.
The gauge-fixed theory has
separate BRST symmetries in the holomorphic and the anti-holomorphic sectors.
For the left-right symmetric case, the quantized chiral gauged WZW theory
is exactly equivalent to the quantized vector gauged WZW theory.
However, the chiral gauged WZW theories also allow us to study the left-right
asymmetric cases, which are useful for heterotic types of string theories.

In Sec. 4, we study the primary fields and their correlation functions
in the gauged WZW theory.  We consider
these fields as the primary fields in the original (ungauged)
WZW theory dressed in the ``clouds''
of the gauge fields.  The gauge fields tend to screen the part of sigma field
that belongs to the gauge group $H_{L,R}$.
This picture suggests that the chiral
gauged WZW theory is in fact the $G/H$ coset theory in conformal field
theory(CFT)\refs{\BPZ,\GKOref}, as was originally proposed by Schnitzer,
Karabali and others for the vector gauged WZW theory\refs{\Schone,\GK}.
When $G$ and $H$ are both simple, it will be shown that
this is indeed the case by an examination of the conformal blocks.
When $G$ is semi-simple, some slight modification has to be made.
To be more explicit, we will discuss in some detail this connection to
coset theory for two cases: (1) $G=SU(2)_k\otimes SU(2)_l$ and $H=SU(2)$.
(2) $G=SU(2)$ and $H=U(1)$.  Recently it was
observed that the $Z_k$ parafermion theory\ZFref (\ie\ the coset
$SU(2)_k/U(1)$)
may play a crucial role in the construction of string theories that have
critical spacetime dimension lower than ten\AT .  These new string theories
have fractional supersymmetry on the world sheet.  An understanding of such
symmetries at the classical action level will be most useful.
In fact it is the attempt to
find a suitable classical action for the parafermion theory that leads to the
present analysis.  Also the relation of gauged WZW theories and
coset models has recently received renewed interests\refs{\EWthree,\BARS} .

In Sec. 5, a couple of explicit examples are presented to illustrate some
properties of chiral gauge theories.
A number of appendices are included to make the paper self-contained.  They
essentially review the various derivations of the results that are needed for
the main text of the paper.  In Appendix A, the evaluation of the non-abelian
anomaly is reviewed. Since it takes no extra effort,
this derivation is presented for the arbitrary (even) dimension case.
In Appendix B, the derivation of the general coupling of the WZW
action to gauge fields with a given anomaly is reviewed.
The evaluation of the determinant used in Sec. 3 is given in Appendix C.

\newsec{\bf Two Different Types of Gauged  WZW Theories}

In general, the coupling of the non-linear sigma field $\phi$ in
the WZW theory to external gauge fields has non-abelian anomalies.
To construct gauged WZW theories (where gauge fields are dynamical),
gauge invariance must be maintained, \ie\ the anomaly must vanish.
This means only special gauge couplings can be introduced, \eg\
vector gauge coupling.  In this section, we shall show that, in two
dimensions in the conformal limit, there exists another type
of gauged WZW theory which is also gauge invariant, \ie\ anomaly-free.

\subsec{\bf The Anomaly in the Effective Theory}

Historically the anomaly first arose in the coupling of  gauge fields
to chiral fermions\Jackiw .
 Let us consider the action
\eqn\Faction
{ S^F=\int d^D x \left( \APsi_R (i\Dirac{\partial}+{\Dirac{A}}^R) \Psi_R
 +\APsi_L (i\Dirac{\partial}+{\Dirac{A}}^L) \Psi_L \right)        }
where the matrix-valued $A^L_{\mu}$ and $A^R_{\mu}$ are  external gauge
fields and
$\Psi_L$ and $\Psi_R$ are multiplets of chiral fermions.  Here we have
suppressed all other fields and their fermionic couplings which are
anomaly-free. (An example will be QCD, where vector gluons are present
and couple to quarks; in this case, $A^{L,R}$ couple to flavor currents.)
The effective action, $W_{\rm eff}^F$,
 of the above action  is obtained through the path integral
\eqn\non{ \exp[-W_{\rm eff}^F(\Apair)]\equiv \int {\cal D}\APsi_R
{\cal D}\Psi_R {\cal D}\APsi_L {\cal D}\Psi_L  \exp(-S^F)~. }
The infinitesimal gauge transformations are defined as
\eqn\chiralagain{\eqalign
{\delta_{\vraised_L} \Psi_L(x)&=v_L(x) \Psi_L(x)~,~~~\delta_{\vraised_R}
              \Psi_R(x)=v_R(x) \Psi_R(x)\cr
\delta A_{\mu}^L &=- \partial_{\mu}\vraised_L+[\vraised_L , A_{\mu}^L]\cr
\delta A_{\mu}^R &=-\partial_{\mu}\vraised_R +[\vraised_R , A_{\mu}^R]\cr }}
where $v_L$ and $v_R$ are matrix-valued and determine infinitesimally the
gauge transformation.
Hence the classical fermion action, $S^F$, is  gauge invariant because
of the minimal couplings in Eq.\Faction .
However, the fermion integration measure may not be invariant under the
 gauge transformations, giving rise to the anomaly in the
effective action\Fuji .  We can express the anomaly in the following
form:
\eqn\anomaly{\eqalign{
 \var W^F_{\rm eff}(\Apair)
&\equiv W^F_{\rm eff}(A^L+\delta A^L,A^R+\delta A^R)- W^F_{\rm eff}(\Apair)\cr
&\equiv c_n \int d^Dx~ \omega^1_{2n}(\Apair;\vpair)\cr }}
where the space-time dimension is $D=2n$ and $\omega^1_{2n}$ is the
non-abelian anomaly, which is a $2n$-form and linear in $v_L$ or in $v_R$.
The constant $c_n$ can be determined through explicit one-loop calculations.
Here $A^{L,R}$ are matrix-valued one forms, \ie , $A^{L,R}\equiv
A^{L,R}_{\mu ,i}{\lambda}^i \df x^{\mu}$ where $\lambda^i$ are group matrices.

The infinitesimal gauge transformation
satisfies the commutation relation
\eqn\anomalyt{
\delta_{v_L,v_R} \delta_{v_L',v_R'}-\delta_{v_L',v_R'} \delta_{v_L,v_R}
=\delta_{[v_L,v_L'],[v_R,v_R']}~.}
This implies that the non-abelian anomaly
must satisfy the following integrability condition,
\eqn\WZcon{\eqalign{
\delta_{v_L,v_R} \int& \omega^1_{2n}(\Apair;v_L',v_R') -
\delta_{v_L',v_R'} \int \omega^1_{2n}(\Apair;v_L,v_R)\cr
&=\int \omega^1_{2n}(\Apair; [v_L,v_L'],[v_R,v_R'])~.  }}
This is the WZ consistency condition\WZref\ and will be used later.
The explicit expression for $\omega^1_{2n}$ is well-known\Zumino ;
its derivation is reviewed in  Appendix A.

On the other hand, the WZW theory can be considered as a low-energy effective
theory of Eq.\Faction\ which must incorporate
the same anomaly effect\WZref . (In the QCD example mentioned
above, we are interested in the effective theory of pions, where the anomaly
plays a crucial role in the $\pi^0$ decay.)
The WZW theory in $2n$-dimensions  is formally defined by
\eqn\Generalwzw{S^B(\phi)={1\over{4\lambda^2}}
      \int_{S^{2n}} d^{2n}x{\rm Tr}\left(
\partial_{\mu}\phi^{-1} \partial^{\mu}\phi\right) - C_n \int_{B^{2n+1}}
{\rm Tr}\left(\df\varphi~ \varphi^{-1} \right)^{2n+1} }
where $B^{2n+1}$ is the ($2n+1$)-dimensional extension of the $2n$-dimensional
space $S^{2n}$, with $S^{2n}$ as its boundary.
$\phi(x)$ is the non-linear sigma field which maps from  $S^{2n}$ to the group
manifold, and $\varphi(x,t)$ is the $(2n+1)$-dimensional extension of $\phi(x)$
whose value on the boundary  $S^{2n}$ is equal to $\phi(x)$.
 $\df$ is the exterior derivative and $(\df\phi
\phi^{-1})^{2n+1}$ is the wedge product of $(2n+1)$ one-forms.
The second term is called the WZ term and will be denoted  by
$\Gamma(\phi)$ in the following.
The effective action of the WZW theory with external gauge fields coupled
to $\phi$ is defined as
\eqn\effective{\exp\left(-W^B_{\rm eff}(\Apair)\right)\equiv\int {\cal D}\phi
                       \exp\left(-S^B(\phi,\Apair)\right)}
where $S^B\left(\phi,\Apair\right)$ is the  WZW action coupled to external
gauge fields.  The gauge coupling to $\phi$ in the first term of
Eq.\Generalwzw\ can be easily incorporated via the minimal
coupling, \ie\ $\partial_{\mu}\phi$ is replaced by the covariant derivative
${\rm D}_{\mu}\phi$ defined as
\eqn\eenj{ {\rm D}_{\mu}\phi \equiv \partial_{\mu}\phi + A_{\mu}^L \phi
-\phi A^R_{\mu}. }
The term ${\rm Tr}\left({\rm D}^{\mu}\phi^{-1}{\rm D}_{\mu}\phi \right)$
is invariant under the gauge transformations
\eqn\chiral{\eqalign{
\phi &\rightarrow g_L(x)~\phi~ g_R^{-1}(x)~\cr
A_{\mu}^L(x)&\rightarrow g_L(x)A_{\mu}^L(x)g^{-1}_L(x)-\partial_{\mu}g_L(x)
\cdot g^{-1}_L\cr
A_{\mu}^R(x)&\rightarrow g_R(x)A_{\mu}^R(x)g^{-1}_R(x)-\partial_{\mu}g_R(x)
\cdot g^{-1}_R\cr}    }
under which ${\rm D}_{\mu}\phi$ behaves as
${\rm D}_{\mu}\phi \rightarrow g_L~  {\rm D}_{\mu}\phi~ g_R^{-1}$.
The infinitesimal version of Eq.\chiral\ is approximated by $g_{L,R}(x) \sim 1+
   \vraised_{L,R}(x)$.  The infinitesimal transformations of $A^{L,R}$ are
the same as those in Eq.\chiralagain\ and that of the sigma field is
\eqn\small{\delta \phi = \vraised_L\phi - \phi \vraised_R ~.}
Under these  gauge transformations, we demand the sigma field's
effective action to have the same anomaly as in Eq.\anomaly ,
\eqn\gaug{\eqalign{ \var W_{\rm eff}^B(\Apair)&=W^B_{\rm eff}\left(
A^L+\delta A^L, A^R+\delta A^R\right)-W^B_{\rm eff}\left(A^L,A^R\right)\cr
  &=c_{n}\int \omega^1_{2n}(\Apair ;\vpair)~. \cr}}
Since the sigma field $\phi$'s path integral measure is gauge invariant, it
follows from Eq.\gaug\ that the classical action, $S^B(\phi,\Apair)$, has
to satisfy
\eqn\generaleq{\var S^B\left(\Apair,\phi\right)=c_n \int \omega^1_{2n}
   (\Apair;\vpair)~.}
Since the first term in $S^B$ is gauge invariant, the anomaly only comes
from the gauged  WZ term.

In the gauged WZW theory,  the   gauge fields become  dynamical
gauge fields via the  introduction of  the kinetic term,
${\rm Tr}(F^2)$.  In the path integral formalism, the gauge field's
configurations are summed over.
Since we would like to preserve the Ward identities,  gauge invariance
of these gauge fields
must be maintained, which is equivalent to  choosing
a specific gauge coupling such that the anomaly $\omega_{2n}^1$ is absent.
This can be achieved in the following way.  First, the
 anomaly $\omega_{2n}^1(\Apair ;\vpair)$ must be determined.
Next, starting from the explicit expression of $\omega_{2n}^1$,
the gauged WZW action $S^B\left(\Apair,\phi\right)$ in Eq.\generaleq\
can be solved.  Finally, among the general solutions of \generaleq , we are
interested in the special gauged WZW actions which are anomaly-free, \ie\ the
corresponding $\omega^1_{2n}=0$.  It means this subset of
$S^B\left(\Apair,\phi\right)$ are gauge invariant.

{} From now on, we are interested only in the 2-dimensional gauged WZW theory
in the conformal limit,  where the ${\rm Tr}(F^2)$ term drops out. (Since
$A^{L,R}_{\mu}$ is of dimension one, the dimension of ${\rm Tr}(F^2)$ is four.
So the ${\rm Tr}(F^2)$ term drops out in the conformal limit.)
The 2-dimensional WZW theory in Eq.\Generalwzw\ can be expressed
as\foot{In the following the superscript ``B'' or ``F''
which distinguishes the boson from the fermion will be neglected.}
\eqn\wzw{S(\phi)={1\over{4\lambda^2}}\int_{S^2}
d^2x{\rm Tr}\left(\partial_{\mu}
\phi^{-1} \partial^{\mu}\phi\right) - {k\over{24\pi}}\int_{B^3} {\rm Tr}
\left(\df\varphi ~ \varphi^{-1} \right)^3~. }
This action has a conformal limit, \ie\ the infrared fixed point\EWtwo ,
specified by $\lambda^2=4\pi/k$.  At this fixed point it is convenient to
use the Euclidean version of the light-cone coordinate
\eqn\eq{z={x_1+ix_2\over{\sqrt 2}}~~~~~ {\rm and}~~~~~
{\bar z}={x_1-ix_2\over{\sqrt 2}}~.}
The WZW action is
invariant under the ``local'' \KM (KM) transformation
denoted by $G_L(z)\otimes G_R({\bar z})$,
\eqn\gltr{ \phi(z,{\bar z}) \rightarrow \Omega_L(z)~\phi(z,{\bar z})~  {}
\Omega_R^{-1}({\bar z})~.}
Note that even though this transformation is ``local'', no gauge fields
have to be introduced to preserve the invariance of the action $S(\phi)$.
This is a special property in two dimensions in the conformal limit.
The symmetry generators of the above transformation are the affine KM
  currents,
whose left currents, $J^{a}(z)t^a=-1/2~k(\partial_z\phi)\phi^{-1}$, and
the right ones, ${\bar J}^{a}({\bar z})t^a=-1/2~k\phi^{-1}(\partial_z
\phi)$ are independent from each  other\refs{\EWtwo,\KZref}.  $t^a$
in these equations are matrices for the group representation and
$k$ is called the level in the affine KM algebra\Kacref. The WZW action in
the conformal limit is denoted by $kI(\phi)$ in the rest of the paper.

Let us make a small digression here.
It is interesting to recall that\KZref , for example, $G_L=G_R=SU(2)$,
\eqn\eq{ {\rm Tr}\left(\partial_{\mu}\phi^{-1} \partial^{\mu}\phi\right)
\sim {\cal G}^{(k)}(z)~ {\bar {\cal G}}^{(k)}({\bar z}) }
where ${\cal G}^{(k)}(z)$ is the fractional supercurrent with dimension
${k+4\over{k+2}}$ (for $k \geq 2$).  So the $SU(2)$ WZW theory
can be expressed as
the WZW theory in the conformal limit perturbed by the current
${\cal G}^{(k)}(z)$,
\eqn\eq{
S(\phi)=kI(\phi)+ \alpha(\lambda)\int d^2z~{\cal G}^{(k)}(z)~
{\bar {\cal G}}^{(k)}({\bar z}) }
where $\alpha(\lambda)$ is a small parameter.  This fractional supercurrent
extends the Virasoro algebra\KMQref\ : ${\cal G}^{(k)}(z)$ and the
stress energy-momentum tensor $T(z)$ form a non-local fractional
superconformal algebra\AGT , which is the basis for fractional
superstring\AT .
(Note that for $k=2$, this is simply the supersymmetric case.)

We have shown that the first term in Eq.\wzw\ can be substituted by  the
gauge invariant expression, ${\rm Tr}\left({\rm D}^{\mu}
\phi^{-1}{\rm D}_{\mu}\phi \right)$, where the covariant derivative
  ${\rm D}_{\mu}\phi$  was  defined in Eq.\eenj .
Therefore, the solution to Eq.\generaleq\ becomes how to determine  the gauge
couplings in the WZ term, $\Gamma(A^L,A^R,\phi)$, \ie\
how to solve
\eqn\gaueq{\var \Gamma\left(\Apair,\phi\right)={k\over{8\pi}}
       \int \omega^1_{2} (\Apair;\vpair)}
where the constant $c_2$ is $k/8\pi$.  However, because of the topological
nature of the WZ term, the gauge couplings in $\Gamma\left(\Apair,\phi\right)$
cannot be introduced simply via the minimal coupling.

Here the anomaly $\omega^1_2(\Apair ;\vpair)$ is known\Zumino :
\eqn\LRano{ \omega^1_{2}(\Apair;\vpair)= {\rm Tr}~(\df v_R~ A^R)
        - {\rm Tr}~(\df v_L~ A^L)~.}
(See Appendix A for a  derivation.)  Given the non-abelian anomaly
$\omega^1_2$ above, Eq.\gaueq\ for the gauged WZ term can be  solved
explicitly, as was done in Ref.\KT .  The basic idea
behind the construction of $\Gamma\left(\Apair,\phi\right)$ is that  the WZ
consistency condition (\ie\ Eq.\WZcon ) allows
the integration of Eq.\gaueq\ along a path in the direction which can be
thought of as the extra dimension needed for the
extension from $S^2$ to $B^3$. The appropriate integration path should
have $\Gamma(A^L,A^R,\phi)$ on the final end point, starting from an explicit
function as the initial point.  We review the detailed derivation in
Appendix B and  give the final answer here,
\eqn\gamans{ \eqalign{ \Gamma(A^L,A^R,\phi)&={-k\over{24\pi}}
\int_{B^3} {\rm Tr} \left(\varphi^{-1}\df\varphi \right)^3 \cr
   &~~~~~~+{k\over{8\pi}}\int_{S^2} {\rm Tr}
\left( A^R\phi^{-1}\df\phi - \df\phi \phi^{-1}A^L +A^R\phi^{-1}A^L\phi
\right)~.  \cr }}
This simply means that the gauged WZ term, $\Gamma(A^L,A^R,\phi)$, takes care
of all the anomaly structure that we require the gauged WZW action
$S(A^L,A^R,\phi)$ to satisfy, \ie , Eq.\generaleq .

Finally, combining Eq.\gamans\ with ${\rm Tr}\left({\rm D}_{\mu}\phi^{-1}
{\rm D}^{\mu}\phi\right)$, we  obtain the gauged WZW action:
\eqn\eq{\eqalign{ S(\Apair,\phi)&={k\over{16\pi}}\int d^2x{\rm Tr}\left(
{\rm D}_{\mu}\phi^{-1} {\rm D}^{\mu}\phi\right) + \Gamma(A^L,A^R,\phi)\cr
&={k\over{16\pi}}\int d^2x{\rm Tr}\left(\partial_{\mu}\phi^{-1}\partial^{\mu}
\phi\right) - {k\over{24\pi}}\int_{B^3} {\rm Tr}\left(\varphi^{-1} \df\varphi
\right)^3\cr
&+{k\over{8\pi}}(\epsilon^{\mu\nu}+g^{\mu\nu})\int d^2x{\rm Tr}\left[
A^{R}_{\mu}\phi^{-1}\partial_{\nu}\phi - \partial_{\mu}\phi \phi^{-1}
A^{L}_{\nu} +A^{R}_{\mu}\phi^{-1}A^{L}_{\nu} \phi\right]\cr
&-{k\over{16\pi}}\int d^2x {\rm Tr}\left( A^{L,\mu}A^L_{\mu}+A^{R,\mu}
A^R_{\mu} \right)\cr } }
If we use the light-cone coordinate, $\epsilon^{z{\bar z}}=
-\epsilon^{{\bar z}z}=1=g^{z{\bar z}}=g^{{\bar z}z}$, we can simplify the
above expression as
\eqn\finalone{ \eqalign{S(\Apair,\phi)=&kI(\phi) + {k\over{4\pi}}
\int d^2z{\rm Tr}\left[A^R_z\phi^{-1}\partial_{\bar z}\phi - A^L_{\bar z}
\partial_z\phi \phi^{-1} + A_z^R\phi^{-1}A^L_{\bar z}\phi\right]\cr
&~~~-{k\over{8\pi}}\int d^2z{\rm Tr}\left(A^L_{z}A^{L}_{\bar z}+A^R_{z}
A^{R}_{\bar z} \right)\cr } }
where $I(\phi)$ can be expressed in this coordinate system as
\eqn\eq{ I(\phi)={1\over{8\pi}}\int d^2z {\rm Tr}\left(\partial_z\phi^{-1}
\partial_{\bar z}\phi\right) -{1\over{8\pi}}\int_{B^3} dtd^2z {\rm Tr}\left(
\varphi^{-1}\partial_t \varphi \varphi^{-1}\partial_z \varphi
\varphi^{-1}\partial_{\bar z} \varphi\right). }
The extra dimension is denoted as $t$. The gauge variation of $S(\Apair,\phi)$
is given by Eq.\generaleq :
\eqn\SYMeq{ \delta_{v_L,v_R}S(\Apair,\phi)={k\over{8\pi}} \int d^2z
{\rm Tr}\left( \partial_z v_R A^{R}_{\bar z} - \partial_{\bar z}v_R A_z^R
     -\partial_z v_L A^{L}_{\bar z} + \partial_{\bar z}v_L A_z^L \right) }
Based on Eq.\SYMeq\ we can discuss at least three possible scenarios
in which the action $S(\Apair,\phi)$ given in Eq.\finalone\
 is invariant under the gauge transformation, \ie\ the non-abelian anomaly
on the R.H.S. of Eq.\SYMeq\ vanishes.
\medskip
\noindent{\bf (I.) Vector Gauge Invariance:}
\smallskip
The vector gauge invariance can be obtained if we choose the gauge
fields and the gauge transformation parameters to be
\eqn\eq{\eqalign{  A^{\rm vec}_z(z,{\bar z})= A^R_z(z,{\bar z})\equiv
    & A^L_z(z,{\bar z})~,~~~~
    A^{\rm vec}_{\bar z}(z,{\bar z})=A^R_{\bar z}(z,{\bar z})\equiv
        A^L_{\bar z}(z,{\bar z})\cr
          & v(z,{\bar z}) = v_R(z,{\bar z}) \equiv v_L(z,{\bar z}) ~.\cr }}
Thus, Eq.\finalone , can be written as
\eqn\Asch{\eqalign{ S(A^{\rm vec}_z,A^{\rm vec}_{\bar z},\phi)=
kI(\phi)+{k\over{4\pi}}&\int d^2z {\rm Tr}\Bigl[A^{\rm vec}_z \phi^{-1}
\partial_{\bar z}\phi - A^{\rm vec}_{\bar z}\partial_z\phi \phi^{-1}\cr
&+ A^{\rm vec}_z \phi^{-1} A^{\rm vec}_{\bar z} \phi - A^{\rm vec}_z
A^{\rm vec}_{\bar z} \Bigr]\cr}}
The vector gauge invariance can be verified from Eq.\SYMeq :
\eqn\eq{ \delta_{v,v} S(A^{\rm vec}_z,A^{\rm vec}_{\bar z} ,\phi)=0~.}
This vector gauged action $S(A^{\rm vec}_z,A^{\rm vec}_{\bar z} ,\phi)$
has been extensively studied  in the  literature\refs{\Schone,
\GK,\EWthree}.  We note that the axial gauge coupling can be introduced
instead of the vector gauge coupling.  In that case
\eqn\eq{\eqalign{  A^{\rm a}_z(z,{\bar z})= A^L_z(z,{\bar z})\equiv
     &-A^R_z(z,{\bar z})~,~~~~
    A^{\rm a}_{\bar z}(z,{\bar z})=A^L_{\bar z}(z,{\bar z})\equiv
        -A^R_{\bar z}(z,{\bar z})\cr
          & v(z,{\bar z}) = v_L(z,{\bar z}) \equiv -v_R(z,{\bar z}) ~.\cr }}

\medskip
\noindent{\bf (II.) Local Ka$\check {\rm \bf c}$-Moody Invariance:}
\smallskip
It is instructive to see how the local KM symmetry, Eq.\gltr , emerges
from our formalism.  Let us choose
\eqn\fthree{\eqalign{ A^L_{\bar z}&\equiv A^R_{z}\equiv 0\cr
         v_L\equiv v_L({z}) ~~~~~~&{\rm and}~~~v_R\equiv v_R({\bar z})\cr}}
so that the gauge fields become
\eqn\eq{      A^L_{\mu}=(A^L_z(z,{\bar z}), 0)~~~{\rm and}~~~
      A^R_{\mu}=(0, A^R_{\bar z}(z,{\bar z}))~.}
In this case, the gauge fields, $A^L_z$ and $A^R_{\bar z}$ decouple from
the $\phi$ field in $S(A^L_z,A^R_{\bar z},\phi)$, \ie\
$S(A^L_z,A^R_{\bar z},\phi)=kI(\phi)$.  Therefore we
recover the original WZW action.  Under the gauge transformation
\eqn\eq{ \delta_{v_L({z}),v_R({\bar z})} S(A^L_{z},A^R_{\bar z},\phi)=
\delta_{v_L({z}),v_R({\bar z})}~kI(\phi)= 0  }
This gauge symmetry is simply the infinitesimal version of the local
transformation in Eq.\gltr , \ie\ $\phi(z,{\bar z}) \rightarrow
\Omega_L(z)\phi(z,{\bar z}) \Omega_R^{-1}({\bar z})$.
Usually the local gauge symmetry implies the introduction of gauge fields.
Hence the local symmetry, Eq.\gltr , of $kI(\phi)$ without gauge fields is
somewhat mysterious.  From our point of view, there are gauge fields $A^L_z,
A^R_{\bar z}$ associated with this local KM symmetry.  It just so happens
that these gauge fields simply decouple from the action $S(A^L_z,
A^R_{\bar z},\phi)$ in Eq.\finalone .

\bigskip
\noindent{\bf (III.) Chiral Gauge Invariance:}
\smallskip

Let us consider another case:
\eqn\eq{\eqalign{ A^L_z&\equiv A^R_{\bar z}\equiv 0\cr
         v_L&=v_L({\bar z}) ~~~~~~{\rm and}~~~v_R=v_R(z)\cr}}
so that  the gauge fields become
\eqn\eq{ A^L_{\mu}=(0, A^L_{\bar z}(z,{\bar z}))~~~{\rm and}~~~
      A^R_{\mu}=(A^R_{z}(z,{\bar z}), 0)~.}
Thus, Eq.\finalone\ becomes
\eqn\ftwo{ S(A^L_{\bar z},A^R_z,\phi) = kI(\phi) + {k\over{4\pi}}
\int d^2z{\rm Tr}\left[A^R_z\phi^{-1}\partial_{\bar z}\phi - A^L_{\bar z}
\partial_z\phi \phi^{-1} +A^R_z\phi^{-1}A_{\bar z}^L\phi \right] }
which will be denoted as the chiral gauged WZW action.  It has the gauge
symmetry following from Eq.\SYMeq\
\eqn\scenone{\delta_{v_L({\bar z}),v_R(z)} S(A^L_{\bar z},A^R_z,\phi)= 0}
This gauge symmetry is new and exists only in 2-dimensions.  It
has the advantage that the left
gauge degrees of freedom are separated from the right ones, \ie\
 $v_L \not= v_R$.  Both the gauge fields,
left-handed and right-handed, have only one component,
$A^L_{\bar z}(z,{\bar z})$ and  $A^R_{z}(z,{\bar z})$ respectively.
In general the chiral gauge invariant action, Eq.\ftwo\ is
different from Eq.\Asch .
The quantization of this chiral gauged WZW theory will be explored
in the next section.

It is worthwhile to mention that  the quantization of the gauge fields
with the vector gauge invariant action, Eq.\Asch\ has been studied extensively
in the literature\refs{\Schone ,\GK}.
However, we shall see that the connection between the coset theory
in the CFT and the gauged WZW theory has a natural setting in terms of the
chiral gauged WZW theory because it has explicit left-right
independent gauge fields.
In Table 1 we summarize the  gauge symmetries of the two different types
of gauged WZW models discussed above.
\bigskip
\medskip
\moveleft2pt
\vbox{\halign{\hfil\footnotefont#\hfil&\quad \hfil\footnotefont#\hfil&\quad
                                               \hfil\footnotefont#\hfil\cr
\figufont {\phantom{symmetry}}&{\bf Chiral Gauge Theory}&
                                  {\bf Vector Gauge Theory}\cr
{\bf Action}&{Eq.\ftwo\  }&  {Eq.\Asch\ }\cr

{\bf Gauge fields}&{$A^L_{\mu}=(0~,~A^L_{{\bar z}}(z,{\bar z}))$}&
   {$A^{\rm vec}_{\mu}=
 \left(A^{\rm vec}_z(z,{\bar z})~,~A^{\rm vec}_{\bar z}(z,{\bar z})\right)$}\cr

{\phantom{symmetry}}&{$A_{\mu}^R=(A^{R}_{z}(z,{\bar z})~,0)$}&
                                               {\phantom{symmetry}}\cr
{\bf Gauge symmetry}&{$\delta\phi=v_L({\bar z})~\phi-\phi ~ v_R(z)$}&
{$\delta\phi=v(z,{\bar z})~\phi-\phi ~v(z,{\bar z})$}\cr
{\phantom{symmetry}}&{$\delta A^L_{\bar z}=-\partial_{\bar z}v_L({\bar z})+
    [v_L,~A^L_{\bar z}]$}&{$\delta A^{\rm vec}_{\bar z}=-\partial_{\bar z}
    v(z,{\bar z})+ [v,~A^{\rm vec}_{\bar z}]$}\cr
{\phantom{symmetry}}&{$\delta A^R_{z}=-\partial_{z}v_R({ z})+
    [v_R,~A^R_{z}]$}&{$\delta A^{\rm vec}_{z}=-\partial_{z}
    v(z,{\bar z})+ [v,~A^{\rm vec}_{z}]$}\cr }}
\bigskip
\centerline{\vbox{   \hsize=3.0in
\item{\bf Table 1.}
Chiral and vector gauged WZW theories.   }}

\bigskip

It will  be shown in Sec. 3 that when the left gauge group is the same as
the right one in the chiral gauge theory, \ie\
  $H_L=H_R$, the quantized chiral gauge theory is
the same as the quantized vector gauge theory.

\newsec{\bf Quantizations of the Chiral Gauged WZW Theory}

The quantization procedure for the chiral gauged WZW theory is very similar
to that for the vector gauged WZW theory\Schone .
Here we shall give a self-contained presentation.

 {}  Formally the partition function  is defined by
\eqn\parti{ {\cal Z} = \int {\cal D} A^R_z {\cal D}A_{\bar z}^L {\cal D}\phi
\exp \left( -S (A^L_{\bar z},A_z^R,\phi) \right) }
where $S(A^L_{\bar z},A_z^R,\phi)$ is given by Eq.\ftwo .
The gauge fields $A_{\bar z}^L$ and $A_z^R$ belong to the adjoint
representations of gauge groups $H_L$ and $H_R$ which are subgroups
of $G_L$ and
$G_R$ respectively.  $A_{\bar z}^L(z,{\bar z})$ and $A_z^R(z,{\bar z})$ can
be parametrized in terms of $h(z,{\bar z})$ and $\widetilde h(z,{\bar z})$ by
\eqn\cha{ A^L_{\bar z} = h^{-1}\partial_{\bar z} h ~~~~~~~~~
 {\rm and}~~~~~~~~A^R_{z}=-\partial_{z}{\widetilde h}{\widetilde h}^{-1}}
These parametrizations simplify the action $S(\Apair,\phi)$ and
facilitate the separation of the gauge
degrees of freedom from the physical degrees of freedom.
By exploiting the Polyakov-Wiegmann formula\PWREF\
\eqn\PW{ I(g\phi) = I(g) +I(\phi) -{1\over{4\pi}}\int d^2x {\rm Tr}
\left(g^{-1}\partial_{\bar z} g~ \partial_{ z}\phi\cdot \phi^{-1}\right)~, }
$S(A^L_{\bar z},A_z^R,\phi)$ can be rewritten in terms of $h, {\widetilde h}$
and ${\widehat \phi}\equiv h \phi{\widetilde h}$, \ie\
$S(A^L_{\bar z},A_z^R,\phi)= S(h,{\widetilde h},{\widehat \phi})$ where
\eqn\chang{S(h,{\widetilde h},{\widehat \phi})=kI({\widehat \phi}) -
k I(h) - kI({\widetilde h})}

The  gauge transformations on $h$, $\widetilde h$ and ${\widehat \phi}$ can
be found from the transformations on $A^L_{{\bar z}}$, $A^R_{z}$ and $\phi$,
\eqn\threeseven{\eqalign{ A^L_{\bar z} &\rightarrow g_L({\bar z})A^L_{\bar z}
      g_L^{-1}({\bar z})-\partial_{\bar z}g_L({\bar z})\cdot g_L^{-1}({\bar z})
=(h~g_L^{-1})^{-1} \partial_{\bar z}(h~g_L^{-1})\cr
&\Rightarrow h\rightarrow h'\equiv h\cdot g_L^{-1}({\bar z})\cr
 A^R_{ z} &\rightarrow g_R({z})A^R_{z}
      g_R^{-1}({z})-\partial_{z}g_R ({z})\cdot
       g_R^{-1}({z})
= - \partial_{z}(g_R~ {\widetilde h} )\cdot (g_R~ {\widetilde h})^{-1}\cr
&\Rightarrow {\widetilde h}\rightarrow {\widetilde h}'\equiv
{g_R}({z}) \cdot  {\widetilde h} \cr
\phi &\rightarrow g_L({\bar z})~ \phi ~g_R^{-1}({z})~~~~~~~
\Rightarrow {\widehat \phi} \rightarrow {\widehat \phi}~,\cr }}
where $g_L({\bar z})$ and $g_R(z)$ are the finite gauge
transformations determined by $v_L({\bar z})$ and $v_R(z)$ respectively.
  {} From Eq.\cha , we see that $A_{\bar z}^L$ is invariant under the
transformation
$h(z,{\bar z})\rightarrow U_L(z) h(z,{\bar z})$.  Similarly,
$A^R_z$ is invariant under the transformation ${\widetilde h}(z,{\bar z})
\rightarrow {\widetilde h}(z,{\bar z}) U_R({\bar z})$.  These are half of the
affine KM symmetry.  Hence the action
Eq.\chang\ is invariant under
\eqn\EQZ{ \eqalign{
h &\rightarrow U_L(z)~h~g_L^{-1}({\bar z}),~~~~{\widetilde h} \rightarrow
g_R({z})~{\widetilde h}~U_R({\bar z})\cr
{\widehat \phi}&=h~\phi~{\widetilde h} \rightarrow
U_L(z)~{\widehat \phi}~U_R({\bar z})\cr }}
Following standard procedures, we shall remove the gauge volume in ${\cal Z}$
in Eq.\parti .  In the new variables $(h, {\widetilde h}, {\widehat \phi})$,
we must also factor out the half-KM symmetry volume (due to the symmetries,
$U_L(z)$ and $U_R({\bar z})$) in the partition function.  This can be achieved
by picking out particular $h_{\rm p}(z,{\bar z})$ and ${\widetilde h}_{\rm p}
(z,{\bar z})$ from their respective gauge orbits and ``half-KM orbits''.  We
shall refer to this procedure loosely as gauge-fixing.

The Jacobian for the change of variable Eq.\cha\ can be calculated from
\eqn\eqA{  {\cal D}A^L_{\bar z}{\cal D}A^R_{z}{\cal D}\phi= J
{\cal D}h~{\cal D}{\widetilde h}~{\cal D}{\widehat \phi} }
where $J$ is the determinant of an upper triangle matrix.
The variation of $A^{L,R}$ can
be determined by that of $h,~{\widetilde h}$,
\eqn\eq{\eqalign{ \delta A^L_{\bar z}&= \delta(h^{-1})\partial_{\bar z} h +
 h^{-1}\partial_{\bar z}(\delta h)\equiv {\rm D}^L_{\bar z}
(\delta h^{-1}\cdot h) \cr
\delta A^R_{z}&= -\partial_z(\delta {\widetilde h}){\widetilde h}^{-1} -
 \partial_z {\widetilde h}~ \delta ({\widetilde h}^{-1}) \equiv {\rm D}^R_z
     (\delta {\widetilde h}\cdot {\widetilde h}^{-1}) \cr  }}
where ${\rm D}_{\mu}^{L,R}$ is the covariant derivative, ${\rm D}^{L,R}
 _{\mu}\equiv -\partial_{\mu} - [A^{L,R}_{\mu},~~~]$.
Therefore the Jacobian is
\eqn\Jch{ J=det({\rm D}^L_{\bar z})~det({\rm D}^R_{z})~.}
These two determinants are evaluated in Appendix C and are equal to
\eqn\temdet{\eqalign{
\left(det~{\rm D}^L_{\bar z}\right)&=\exp\left[2C_2(H_L)I(h)
\right] \cdot \left(\prod_{\a ,\b =1}^{dim~H_L}det~\delta^{\a\b}~
\partial_z\right)\cr
\left(det~{\rm D}^R_{\bar z}\right)&=\exp\left[
2C_2(H_R)I({\widetilde h}) \right]\cdot \left(
  \prod_{\a ,\b =1}^{dim~H_R}det~\delta^{\a\b}~\partial_{\bar z} \right)~.\cr}}
where $C_2$ is the dual Coxeter number, defined by the structure
constant of the gauge group,
$\sum_{\b \c}f^{a \b \c}f^{\a \b \c}=-C_2\delta^{a\a}$.

Finally we can combine Eq.\parti , \cha , \chang , \Jch ,
  \temdet\ and exponentiate the determinants $\left(det~\delta^{\a\b}~
\partial_{\bar z}\right)$ and $\left(det~\delta^{\a\b}~\partial_{z}\right)$
 to rewrite the  full quantum partition function as
\eqn\quan{\eqalign{ &{\cal Z}_{\rm gauged-fixed} = \int
{\cal D}h_{\rm p} {\cal D}{\widetilde h}_{\rm p} {\cal D}{\widehat\phi}
\exp\left(-kI(\widehat\phi)\right)\cr
&~~~\left[ \exp \Bigl( \left[k+2C_2(H_L)\right] I(h_{\rm p}) \Bigr)
\left(\prod_{\a =1}^{dim H_L} {\cal D}b_z^{\a}{\cal D}c^{\a} \exp\left(
-\int d^2z b_z^{\a}{\bar \partial}c^{\a}\right) \right)\right]\cr
&~~~\left[\exp\left( \left[k+2C_2(H_R)\right] I({\widetilde h}_{\rm p})\right)
\left(\prod_{\a =1}^{dim H_R} {\cal D}{\bar b}_{\bar z}^{\a}{\cal D}
{\bar c}^{\a}  \exp\left(
-\int d^2z {\bar b}_{\bar z}^{\a} {\partial}{\bar c}^{\a}\right)\right)
\right]\cr } }
where $b_z^{\a}, c^{\a}$ (and ${\bar b}^{\a}_{\bar z}, {\bar c}^{\a}$) are
spin $(1,0)$ ghosts introduced for each generator of the gauge group $H_L$
(and $H_R$).   The 2-point function, mode expansions and anticommutations
of $b_z^{\a}, c^{\a}$ are standard:
\eqn\ghoone{\eqalign{ b_z^{\a}(z) c^{\b}(w) = {\delta^{\a\b}\over{z-w}}~~~
 &{\rm and}~~~b_z^{\a}(z) b_z^{\b}(w)=c^{\a}(z) c^{\b}(w)=0\cr
b_z^{\a}(z)=\sum_nz^{-n-1}b^{\a}_{z,n}~~~&{\rm and}
 {}    ~~~c^{\a}(z)=\sum_nz^{-n}c^{\a}_n\cr
\{ b^{\a}_{z,n}~,~~c^{\b}_m\} =\delta^{\a\b}\delta_{n+m}~~~&{\rm and}~~~
\{b^{\a}_{z,n}~,~~b^{\b}_{z,m}\} =\{ c^{\a}_n~,~~c^{\b}_m\} =0~.\cr}}
${\bar b}^{\a}_{\bar z}, {\bar c}^{\a}$ have similar constructions.

As familiar in the quantum gauge theory, the inclusion of the ghost fields
allows us to construct the BRST operator whose nilpotency guarantees
that the gauge symmetry is maintained at the quantum level.
The construction of the BRST operators relies on the remaining symmetries in
the gauge-fixed ${\cal Z}$ in Eq.\quan .
To find out such symmetry we first review the affine KM symmetries in the WZW
actions $I(h_{\rm p})$, $I({\widetilde h}_{\rm p})$ and $I({\widetilde \phi})$.
The KM currents are denoted by $J^{\a}_{h}(z)$,
${\bar J}^{\b}_{\tilde h}({\bar z})$, $J_{\phi}^a(z)$ and
${\bar J}_{\phi}^{a}({\bar z})$ where $a= 1,2,...,dim~G$,
$\alpha = 1,2,...,dim~H_L$ and $\beta = 1,2,...,dim~H_R$.  Those currents
 have the operator product expansions(OPEs)
\eqn\ghotwo{\eqalign{
J_{\phi}^a(z) J_{\phi}^b(w) &= {\delta^{ab}k/2\over{(z-w)^2}}+{f^{abc}
J_{\phi}^c\over{(z-w)}} + {\rm regular~ terms.} \cr
J_{h}^{\a}(z) J_{h}^{\b}(w) &= {\delta^{\a\b}[-k/2-C_2(H_L)]\over{(z-w)^2}}+
{f^{\a\b\c}J_{h}^{\c}\over{(z-w)}} +  {\rm regular~ terms.} \cr } }
where $f^{abc}$ and ${f}^{\alpha\beta\gamma}$ are the structure
constants of the group $G_L$ and the gauge group $H_L$ respectively.
The right-handed currents can be similarly constructed.
The mode expansions in terms of moding operators of these currents are
$J_{\phi}^a(z)=\sum_n z^{n-1} J_{\phi ,-n}^a$ and
${\bar J}_{\phi}^{a}$, $J^{\a}_h$ and ${\bar J}_{\tilde h}^{\alpha}$ have
similar expressions.  The commutation relations between the moding operators
can be obtained through
the above OPEs. The ghost action in Eq.\quan\ also has affine KM symmetries
generated by KM ghost currents
\eqn\eq{J_{\rm gh}^{\a}(z)=-f^{\a\b\c}:b_{z}^{\b}c^{\c}:(z) }
which satisfy affine KM algebra of group $H_L$ at level $2C_2(H_L)$.
So we can define
\eqn\CONtot{\eqalign{
 J_{\rm tot}^a(z)&\equiv J_{\phi}^a(z)~~~~~{\rm for}~~a\in G_L,~~{\rm and}~~
  a\not\in H_L \cr
J_{\rm tot}^{\a}(z)&\equiv J_{\phi}^{\a}(z)+J_h^{\a}(z)+J^{\a}_{\rm gh}(z)~~~~~
{\rm for}~~ \a\in H_L~.\cr}}
Note that the $(z-w)^{-2}$ term is absent in the OPE
\eqn\eq{
J_{\rm tot}^{\a}(z)J_{\rm tot}^{\b}(w)={f^{\a\b\c}J_{\rm tot}^{\c}(w)\over{
z-w}}+... ~~~~~{\rm for}~~\a ,\b~\in H_L}
\ie\ the $\{ J_{\rm tot}^{\a}(z),~~{\rm for}~\a\in H_L\} $ current algebra
is free from affine KM anomaly as required by the gauge invariance.
  Similarly the antiholomorphic KM currents ${\bar J}_{\rm tot}^{\a}
({\bar z})$ can be constructed.
Hence the holomorphic and antiholomorphic BRST operators can be found
\eqn\brstLR{\eqalign{ Q_{\rm BRST}^L&=\oint dz~ \prod_{\a =1}^{dim H_L}
c^{\a}(z) J_{\rm tot}^{\a}(z)\cr
Q_{\rm BRST}^R&=\oint d{\bar z}~\prod_{\alpha =1}^{dim H_R} {\bar c}^{\a}
({\bar z}){\bar J}_{\rm tot}^{\a}({\bar z})\cr }}
These two BRST operators correspond exactly to the gauge symmetry
generated by $\vraised_L({\bar z})$ and $\vraised_R({z})$.
They can also be written in terms of the moding operators
of the currents and the $b, c$ ghost, \eg\
\eqn\eq{
Q^L_{BRST}=\sum^{\infty}_{\infty} c^{\a}_{-n}\left(J_{\phi ,n}^{\a}+
J^{\a}_{h,n}\right) -
{1\over 2}f^{\a\b\c}:c^{\a}_{-n}b^{\b}_{-m}c^{\c}_{n+m}: }
The anticommutation relations between these BRST operators can be easily
checked from Eq.\ghoone , \ghotwo\ and \brstLR ,
\eqn\eq{
\left\{Q_{BRST}^L~,~~Q_{BRST}^L\right\}=\left\{Q_{BRST}^R~,~~Q_{BRST}^R\right\}
=\left\{Q_{BRST}^L~,~~Q_{BRST}^R\right\}=0}
which guarantee the nilpotency of the BRST operators.

There is a special case in which $h_{\rm p}$ and ${\widetilde h}_{\rm p}$
fields decouple from the matter and ghost system.  It can be
seen from Eq.\quan\ that when the level $k=-2C_2(H_L)$, the $h_{\rm p}$ field
decouples from the action and the integration over $h_{\rm p}$ becomes an
irrelevant volume factor which can be factored out from the path integral.
This  can also be seen from the
construction of the BRST operators without including $h_{\rm p}$
in $Q_{BRST}^L$, \ie\
\eqn\Zvon{ Q_{BRST}^L=\oint dz~ \prod_{\a =1}^{dim H_L}
c^{\a}(z) \left[ J_{\phi}^{\a} (z)+ J_{\rm gh}^{\a}(z)\right] }
which satisfies
\eqn\eq{ \left\{Q_{BRST}^L~,~~Q_{BRST}^L\right\}=\left(
      k+2C_2(H_L)\right)\sum^{\infty}_{m=1} m c^{\a}_{-m} c^{\a}_{m}. }
Hence the nilpotency of $Q_{BRST}^L$ without $h_{\rm p}$ is guaranteed if
the level $k=-2C_2(H_L)$; similarly $Q_{BRST}^R$ is nilpotent without
${\widetilde h}_{\rm p}$ if the level $k=-2C_2(H_R)$.
Therefore, $k$ has the {\it critical} value when $k=-2C_2(H_L)=-2C_2(H_R)$
because $h_{\rm p}$ and ${\widetilde h}_{\rm p}$ both decouple from the theory.
Thus the chiral gauged WZW theory is remarkably similar to the vector gauged
WZW theory.  The $Q_{BRST}$ in Eq.\brstLR\ was first considered in Ref.\Schone\
and the special case Eq.\Zvon\ was first considered in Ref.\ZY .

\newsec{\bf Chiral Gauged WZW Theories and Coset Models }

Now we are ready to demonstrate the equivalence of the gauged WZW theories and
the coset models in CFT.
Our analysis is motivated by the work on the two-dimensional
quantum gravity\ddk\ and the starting point is Knizhnik and Zamolodchikov's
formulation of WZW theory.
The primary fields in the chiral gauged WZW theory are
considered as the primary fields in the original (\ie\ ungauged)
 WZW theory dressed in the
``clouds'' of the gauge fields.  By an examination of
the correlation functions (or more precisely the conformal blocks)
of these dressed primary fields, we shall
show that the chiral gauged WZW theory are isomorphic to
the coset theory $G/H$.  This result follows from the fact that the ``gauge
clouds'' surrounding the primary fields of the original ungauged WZW theory
are precisely the primary fields of $H$ but with a time-like metric.  So they
effectively cancel that part of the primary fields of $G$ which belongs to $H$.
For the sake of concreteness, we will consider two examples in some
detail: (1) $G=SU(2)_k$ with $H_L=H_R=U(1)$ and its equivalence to
the $Z_k$ parafermion (\ie\ $SU(2)_k/U(1)$ coset) theory;
(2)  $G=SU(2)_k\otimes SU(2)_l$ with $H=SU(2)$ will be shown to be isomorphic
to the coset theory $\SU$.

As  discussed in the last section, there are affine \KM\ currents
$J_{\phi}^a(z)$ and ${\bar J}_{\phi}^{a}({\bar z})$ in the WZW theory,
whose OPEs are listed in Eq.\ghotwo .
The primary fields $G^{\Lambda ,{\bar \Lambda}}_{\lambda,
{\bar \lambda}}(z,{\bar z})$ of the WZW theory with quantum number $\lambda,
{\bar \lambda}$ in the representations $\Lambda ,{\bar \Lambda}$ satisfy
\eqn\JANfive{\eqalign{
J_{\phi}^a(z)G^{\Lambda ,{\bar \Lambda}}_{\lambda,{\bar \lambda}}(w,{\bar w})
&={(t^a_{\Lambda})_{\lambda,{\lambda}'}G^{\Lambda ,
{\bar \Lambda}}_{{\lambda}',{\bar \lambda}}(w,{\bar w})\over{z-w}}
  + {\rm reg.~terms}\cr
{\bar J}_{\phi}^a({\bar z})G^{\Lambda ,{\bar \Lambda}}_{\lambda,
{\bar \lambda}}(w,{\bar w})&={(t^a_{\bar \Lambda})_{{\bar \lambda},
{\bar \lambda}'}G^{\Lambda ,{\bar \Lambda}}_{{\lambda},{\bar \lambda}'}
(w,{\bar w})\over{{\bar z}-{\bar w}}}   + {\rm reg.~terms}\cr  }}
where $t^a_{\Lambda}$ and $t^a_{\bar \Lambda}$ are the matrices of the
group generators in the representations $\Lambda$ and ${\bar \Lambda}$
respectively.  The Fock space of the WZW theory
is built up by applying the negative modings of
the holomorphic and antiholomorphic currents on
$G^{\Lambda ,{\bar \Lambda}}_{\lambda,{\bar \lambda}}
(w,{\bar w})$.  Because $J_{\phi}^a(z)$'s are
independent from ${\bar J}_{\phi}^a({\bar z})$'s, we can separate
the holomorphic and antiholomorphic parts of the fields in the WZW theory, \ie\
$G^{\Lambda ,{\bar \Lambda}}_{\lambda,{\bar \lambda}}(w,{\bar w})~=
G^{\Lambda}_{\lambda}(w)G^{{\bar \Lambda}}_{\bar \lambda}({\bar w})$
and similarly for all other states in the Fock space.
The holomorphic part of the stress energy-momentum tensor in the WZW theory
is given by the Sugawara form
\eqn\JANtwo{
T_{\phi}(z)={1\over{k+C_2(G)}}\sum_{a=1}^{dim~G}
 :J_{\phi}^a(z)J_{\phi}^a(z):~. }
Therefore the highest weights of $G^{\Lambda ,{\bar \Lambda}}_{\lambda,
{\bar \lambda}}(w,{\bar w})$ can be obtained from Eq.\JANtwo\ and \JANfive\
\eqn\JANsix{\Delta_G^{\Lambda}={C_{\Lambda}(G)\over{k+C_2(G)}} ~~~{\rm and}~~~
     {\bar \Delta}^{\bar\Lambda}_G={C_{\bar \Lambda}(G)\over{k+C_2(G)}} }
where $C_{\Lambda}$ is the quadratic Casimir of the representation $\Lambda$.

In the chiral gauged WZW theory, the currents are given in Eq.\CONtot ,
\eqn\CONtottwo{\eqalign{
 J_{\rm tot}^a(z)&\equiv J_{\phi}^a(z)~~~~~{\rm for}~~a\in G_L,~~{\rm and}~~
  a\not\in H_L \cr
J_{\rm tot}^{\a}(z)&\equiv J_{\phi}^{\a}(z)+J_h^{\a}(z)+J^{\a}_{\rm gh}(z)~~~~~
{\rm for}~~ \a\in H_L~.\cr}}
The OPEs among the currents are listed in Eq.\ghotwo .
The holomorphic stress energy-momentum tensor of the chiral gauged WZW theory
is
\eqn\JANone{ T_{\rm tot}(z)=T_{\phi}(z)+T_{h}(z)+T_{\rm gh}(z) }
where  $T_{h}(z)$ is also given by the Sugawara form
\eqn\JANthree{\eqalign{
T_{h}(z)&={1\over{[-k-2C_2(H_L)]+C_2(H_L)}}\sum_{\a =1}^{dim~H_L}
 :J_{h}^{\a}(z)J_{h}^{\a}(z):\cr
&={-1\over{k+C_2(H_L)}}\sum_{\a =1}^{dim~H_L}:J_{h}^{\a}(z)J_{h}^{\a}(z): }}
where the additional minus sign follows from that in Eq.\quan .
The stress energy-momentum tensor of the ghost, $T_{\rm gh}(z)$, is
\eqn\eq{
T_{\rm gh}(z)=-\sum_{\a=1}^{dim~H_L} b_z^{\a}(z)\partial_z c^{\a}(z)~.}
Therefore the central charge of $T_{\rm tot}(z)$ is precisely that for the
$G/H$ coset model,
\eqn\JANfour{\eqalign{
c_{\rm tot}&={k\cdot dim(G_L)\over{k+C_2(G_L)}}
+{[-k-2C_2(H_L)]\cdot dim(H_L)\over{-k-C_2(G_L)}} -2\cdot dim(H_L)\cr
&={k\cdot dim(G_L)\over{k+C_2(G_L)}} - {k\cdot dim(H_L)\over{k+C_2(H_L)}}
=c_G-c_H\equiv c_{G/H}\cr }}
which was first observed in Ref.\Schone .

Let us recall from the previous section that the physical primary state,
$\vert \Phi(w,{\bar w})>\otimes {\vert 0>}^L_{\rm gh}
\otimes {\vert 0>}^R_{\rm gh}$, in the chiral gauged WZW theory are the BRST
singlet states
\eqn\eq{ Q^{L,R}_{BRST}\left( \vert \Phi(w,{\bar w})> \otimes
{\vert 0>}^L_{\rm gh} \otimes {\vert 0>}^R_{\rm gh} \right) =0 }
where ${\vert 0>}^{L,R}_{\rm gh}$ are the ghost vaccum satisfying, for
$\a = 1,2, ..., dim H_L$,
\eqn\eq{c_n^{\a} {\vert 0>}^L_{\rm gh} =0~~~n\geq 1~,~~~{\rm and}~~~~
b_n^{\a} {\vert 0>}^L_{\rm gh} =0~~~n\geq 0~. }
The state ${\vert 0>}^L_{\rm gh}$ is also annihilated by the zero mode
of the ghost
current, $J_{{\rm gh},0}^{\alpha}{\vert 0>}^L_{\rm gh}=0$.
Similar constraints apply to ${\vert 0>}^R_{\rm gh}$.
Therefore from the
construction of BRST operators, Eq.\CONtot\ and \brstLR , we obtain
\eqn\A{\eqalign{
J_{\rm tot}^{\alpha}(z) \Phi(w,{\bar w})&=\left(J_{\phi}^{\alpha}(z)
+J_{h}^{\alpha}(z)\right) \Phi(w,{\bar w})=
{\rm reg.~terms}\cr
{\bar J}_{\rm tot}^{\beta}({\bar z}) \Phi(w,{\bar w})&=
\left({\bar J}_{\phi}^{\beta}({\bar z})
+{\bar J}_{\tilde h}^{\beta}({\bar z})\right) \Phi(w,{\bar w})=
{\rm reg.~terms} \cr }}
where $\beta = 1,2, ..., dim~H_R$.
The physical primary fields in the chiral gauged WZW theory can be
considered as the WZW primary fields dressed up by a cloud of the gauge
fields, $h_{\rm p}$ and ${\widetilde h}_{\rm p}$,
\eqn\eq{
\Phi^{\Lambda ,{\bar \Lambda}}_{\lambda,{\bar \lambda}}(w,{\bar w})
=F_1(h_{\rm p})(w,{\bar w})\left( G^{\Lambda}_{\lambda}(w)
 G^{{\bar \Lambda}}_{\bar \lambda}({\bar w})\right)
F_2({\widetilde h}_{\rm p})(w,{\bar w}). }
The above factorization property is also justified from the independence
of $J_{\phi}^{\a}(z)$ and $J^{\a}_{h}(z)$ (also ${\bar J}_{\phi}^{\a}
({\bar z})$
 and $J^{\a}_{\tilde h}({\bar z})$).
Since the $h_{\rm p}$ field only appears in the holomorphic currents,
$J_{\rm tot}^{\alpha}(z)$ and $T_{\rm tot}(z)$, and similarly the ${\tilde h}_
{\rm p}$ field only appears in the anti-holomorphic currents,
${\bar J}_{\rm tot}^{\alpha}({\bar z})$ and ${\bar T}_{\rm tot}({\bar z})$,
 it is natural to choose the dressing $F_1(h_{\rm p})$ as a holomorphic
function $F_1(h_{\rm p})(z)$ and the dressing
$F_2({\tilde h}_{\rm p})$ as an anti-holomorphic function
$F_2({\tilde h}_{\rm p})({\bar z})$.  In other words,
Eq.\A\ can be further simplified as
\eqn\JANseven{\eqalign{
\left(J_{\phi}^{\a}(z) +J_{h}^{\a}(z)\right)\Phi^{\Lambda}_{\lambda}(w)&=
{\rm reg.~terms}\cr
\left({\bar J}_{\phi}^{\b}({\bar z}) +J_{\tilde h}^{\b}({\bar z})\right)
{\bar \Phi}^{{\bar \Lambda}}_{\bar \lambda}({\bar w})&={\rm reg.~terms}\cr }}
where
\eqn\JANeight{
\Phi^{\Lambda}_{\lambda}(w)=H^{\Lambda}_{\lambda}(w)\cdot G^{\Lambda}_{\lambda}
(w)~,~~~~{\bar \Phi}^{{\bar \Lambda}}_{\bar \lambda}({\bar w})
={\bar H}^{{\bar \Lambda}}_{\bar \lambda}({\bar w})\cdot {\bar G}^{\bar
\Lambda}_{\bar \lambda}({\bar w})~.}
and $H^{\Lambda}_{\lambda}(w)$ satisfy
\eqn\JANnine{
J_{h}^{\a}(z) H^{\Lambda}_{\lambda}(w)={-(t^{\a}_{\Lambda , H})_{\lambda,
{\lambda}'}H^{\Lambda}_{{\lambda}'}(w)\over{z-w}}
  + {\rm reg.~terms}  }
where $(t^{\a}_{\Lambda , H})_{\lambda,{\lambda}'}=(t^{\a}_{\Lambda})_{
\lambda,{\lambda}'}$ if $\a = 1,2,...,~dim~H_L$.  To illustrate Eq.\JANnine ,
consider $G=SU(3)$, $H=SU(2)$ and $\Lambda =$ the fundamental representation
of $SU(3)$.  The $t^{\a}_{\Lambda}$'s defined in Eq.\JANfive\ are the Gell-Mann
matrices, where $a=1,2,...,8$.  Then the matrices $t^{\a}_{\Lambda , H}$
for $\a=1,2,3$,
can be decomposed into a block-diagonal form in terms of the representation
matrices of $SU(2)$,
\eqn\eq{
\left( t^{\a}_{\Lambda , H}\right) \equiv
\left( t^{\a}_{j=1/2}\right)~\oplus~\left( t^{\a}_{j=0}\right) }
where $\left( t^{\a}_{j}\right)$ are the spin-j representation matrices
of $SU(2)$ with dimension $(2j+1)\times (2j+1)$.  Here $\left(
t^{\a}_{j=1/2}\right)$ are
Pauli matrices and $\left(t^{\a}_{j=0}\right)=0$.
${\bar H}^{{\bar \Lambda}}_{\bar \lambda}({\bar w})$'s satisfy similar
constraints.  The minus sign in front of $\left( t^{\a}_{\Lambda , H}\right)$
indicates that $H^{\Lambda}_{\lambda}(w)$'s are time-like fields.  The highest
weight of $H^{\Lambda}_{\lambda}(w)$ with respect to the stress
energy-momentum, $T_h(z)$,  Eq.\JANthree , is
\eqn\DIE{
T_h(z) H^{\Lambda}_{\lambda}(w) = {-\Delta_{H_L}^{\lambda}\over{(z-w)^2}}
H^{\Lambda}_{\lambda}(w) +...}
where $\Delta_{H_L}^{\lambda}=C_{\lambda}(H_L)/(k+C_2(H_L))$.
$C_{\lambda}(H_L)$ means symbolically the quadratic Casimir of the
representation that $\lambda$ belongs to.
Note that the additional minus sign in Eq.\DIE\ follows from Eq.\JANthree .
Therefore the highest weight of $\Phi^{\Lambda}_{\lambda}(w)$ is
\eqn\JANnineone{
T_{\rm tot}(z) \Phi^{\Lambda}_{\lambda}(w) =
{\Delta_G^{\Lambda}-\Delta_{H_L}^{\lambda}\over{(z-w)^2}}\Phi^{\Lambda}_
{\lambda}(w) +... }
which  is the same as that of the primary fields in the
coset theory $G/H$.  In the above example $G/H=SU(3)/SU(2)$,
$\Delta_{H_L}^{\lambda}={j(j+1)\over{k+2}}$.  Therefore, with respect to $T_h$,
the highest weights of $H^{\Lambda}_{\lambda}(w)$, for $\lambda =1,2$ and
that of $H^{\Lambda}_{\lambda =3}(w)$ are respectively
\eqn\eq{\eqalign{
\Delta_{H_L}^{\lambda =1,2}&={3\over{4(k+2)}}\cr
\Delta_{H_L}^{\lambda =3}&=0~.\cr  }}

The holomorphic part of the correlation functions of these dressed primary
fields can be calculated as
\eqn\JANten{\eqalign{
&<\Phi^{\Lambda_1}_{\lambda_1}(z_1)\Phi^{\Lambda_2}_{\lambda_2}(z_2)...
\Phi^{\Lambda_n}_{\lambda_n}(z_n)>\cr
&=<G^{\Lambda_1}_{\lambda_1}(z_1)G^{\Lambda_2}_{\lambda_2}(z_2)...
G^{\Lambda_n}_{\lambda_n}(z_n)>
 <H^{\Lambda_1}_{\lambda_1}(z_1)H^{\Lambda_2}_{\lambda_2}(z_2)...
H^{\Lambda_n}_{\lambda_n}(z_n)>~.\cr }}
Since this expression is in general a linear combination of conformal blocks,
we may for convenience consider one conformal block at a time.
The primary field $G^{\Lambda}_{\lambda}$ can be decomposed into
$G^{\Lambda}_{\lambda}(z)=g^{\Lambda}_{\lambda}(z){\Omega}^{\Lambda}_{\lambda}
(z)$ where ${\Omega}^{\Lambda}_{\lambda}(z)$'s satisfy
\eqn\eq{\eqalign{
J_{\phi}^{\a}(z) {\Omega}^{\Lambda}_{\lambda}(w)&=
{(t^{\a}_{\Lambda , H})_{\lambda,{\lambda}'}{\Omega}^{\Lambda}_{{\lambda}'}
(w)\over{z-w}}  + {\rm reg.~terms}  \cr
J_{\phi}^{\a}(z) g^{\Lambda}_{\lambda}(w) &= {\rm reg.~terms}
 {} ~~~~~~~~~~~~~~~~~~{\rm for}~\a  = 1,2, ..., dim~H_L\cr  }}
where $(t^{\a}_{\Lambda , H})$ is the same as that in Eq.\JANnine .

Because of the time-like nature of the dressed field $H^{\Lambda}_{\lambda}$
(due to the
additional minus sign in front of $I(h_{\rm p})$ in Eq.\quan ), the second
factor on the right-hand side of Eq.\JANten\ is the reciprocal of the
conformal block of the corresponding fields in the representation of
the  subgroup $H_L$, \ie\
\eqn\eq{
<H^{\Lambda_1}_{\lambda_1}(z_1)H^{\Lambda_2}_{\lambda_2}(z_2)...
H^{\Lambda_n}_{\lambda_n}(z_n)>
=\left\{<{\Omega}^{\Lambda_1}_{\lambda_1}(z_1)
{\Omega}^{\Lambda_2}_{\lambda_2}(z_2)...
{\Omega}^{\Lambda_n}_{\lambda_n}(z_n)>\right\}^{-1}~.}
Therefore Eq.\JANten\ is equal to the conformal block in the coset
theory $G/H$, \ie\
\eqn\JANtwelve{
<\Phi^{\Lambda_1}_{\lambda_1}(z_1)\Phi^{\Lambda_2}_{\lambda_2}(z_2)...
\Phi^{\Lambda_n}_{\lambda_n}(z_n)>
=<g^{\Lambda_1}_{\lambda_1}(z_1)g^{\Lambda_2}_{\lambda_2}(z_2)...
g^{\Lambda_n}_{\lambda_n}(z_n)>}
 The antiholomorphic parts can be similarly calculated.
To obtain the correlation function,
we  have to impose the monodromy invariant and fusion invariant conditions
in combining the holomorphic and the
anti-holomorphic conformal blocks.
For simple theories, consistent correlation functions exist only for
$H_L=H_R$, which is equivalent to the vector gauged WZW theories.
However, in the construction of heterotic type of string models\keith ,
there exists consistent left-right asymmetric correlation functions which
can be possibly provided by the chiral gauged WZW theories.

To illustrate in more detail the general formalism presented above, two
examples are given in the following, \ie\ $Z_k$ parafermion (PF) (\ie\
$SU(2)_k/U(1)$ coset theory) and $SU(2)_k\otimes SU(2)_l/SU(2)_{k+l}$
coset model.

\medskip
\noindent{{\bf (I.)~${\bf G}=SU(2)_k,~{\bf H}=U(1):$}}
\smallskip

For the sake of concreteness, we briefly summarize some results in
 the $SU(2)_k$ WZW theory\KZref .
The holomorphic  $SU(2)_k$ KM current algebra has the following OPEs among
its dimension-one currents:
\eqn\PPr{\eqalign{
J_{\phi}^+(z)J_{\phi}^-(w)&={k\over{(z-w)^2}}+{2J_{\phi}^3(w)\over{z-w}} +
     {\rm reg.~ terms}\cr
J_{\phi}^3(z)J_{\phi}^{\pm}(w)&={\pm J_{\phi}^{\pm}(w)\over{z-w}} +
   {\rm reg.~ terms} \cr
J_{\phi}^3(z)J_{\phi}^3(w)&={k/2\over{(z-w)^2}} + {\rm reg.~ terms} \cr }}
The holomorphic stress energy-momentum tensor is given by the Sugawara form
\eqn\eq{ T_{SU(2)}(z)={1\over{k+2}}\sum_a :J_{\phi}^a(z)J_{\phi}^a(z): }
whose central charge is $c_{SU(2)}=3k/(k+2)$.
The antiholomorphic currents, ${\bar J}_{\phi}^3({\bar z})$ and
${\bar J}_{\phi}^{\pm}({\bar z})$, commute with the holomorphic currents and
obey a similar algebra.  The primary fields in the $SU(2)_k$
WZW theory are denoted by $G^{j,{\bar j}}_{j,{\bar j}}(z,{\bar z})$, satisfying
\eqn\eq{\eqalign{
J^a_{\phi ,n} G^{j,{\bar j}}_{j,{\bar j}}(z,{\bar z})=~&0~=
{\bar J}^a_{\phi ,n}G^{j,{\bar j}}_{j,{\bar j}}(z,{\bar z}),~~~~~~~~
   {\rm for}~n>0 \cr
J_{\phi ,0}^+G^{j,{\bar j}}_{j,{\bar j}}(z,{\bar z})=~&0~={\bar J}_{\phi ,0}^+
G^{j,{\bar j}}_{j,{\bar j}}(z,{\bar z})\cr
J_{\phi ,0}^3G^{j,{\bar j}}_{j,{\bar j}}(z,{\bar z})=j&
G^{j,{\bar j}}_{j,{\bar j}}(z,{\bar z})~,~~~~~~~
{\bar J}^3_{\phi ,0}G^{j,{\bar j}}_{j,{\bar j}}(z,{\bar z})={\bar j}
G^{j,{\bar j}}_{j,{\bar j}}(z,{\bar z})\cr }}
where $J^a_{\phi ,n}$ and ${\bar J}^a_{\phi ,n}$ are the moding operators
defined by $J^a_{\phi}(z)=\sum _nJ^a_{\phi ,n}z^{-n-1}$ and
${\bar J}_{\phi}^a({\bar z})=\sum _n{\bar J}^a_{\phi ,n}
{\bar z}^{-n-1}$.  The Virasoro primary fields, $G^{j,{\bar j}}_{m,{\bar m}}
(z,{\bar z})$, are defined by
\eqn\eq{
G^{j,{\bar j}}_{m,{\bar m}}(z,{\bar z})~=~(J^-_{\phi ,0})^{j-m}
({\bar J}_{\phi ,0}^-)^{{\bar j}
-{\bar m}} G^{j,{\bar j}}_{j,{\bar j}}(z,{\bar z}) }
whose $J_{\phi ,0}^3$ quantum number is $m$ and the highest weight is
$\Delta^j=j(j+1)/(k+2)$.  As discussed above, the primary fields can be
separated into the holomorphic and antiholomorphic parts, \ie\
$G^{j,{\bar j}}_{m,{\bar m}}(z,{\bar z})~=G^j_m(z){\bar G}^{\bar j}_{\bar m}
({\bar z})$.

Now let us consider the chiral gauged WZW theory where $H_L=U(1)$.
The left $U(1)$ gauge field $A^L_{\bar z}(\equiv h^{-1}\partial_{\bar z}h)$
has the action as shown in Eq.\quan\
\eqn\free{
S(h)=-kI(h)={-k\over{16\pi}}\int d^2z \partial_{\mu}h^{-1}
          \partial^{\mu}h={-1\over{8\pi}} \int d^2z \partial_{z}\sigma
          \partial_{\bar z} \sigma}
where the WZ term vanishes and $C_2(U(1))=0$.  In Eq.\free ,
$h$ has been parametrized by
\eqn\eq{ h =\exp\left(-i\sigma(z,{\bar z})/{\sqrt{k}}\right).}
Eq.\free\ is the action of a free time-like boson with the equation of motion,
$\partial_z \partial_{\bar z}\sigma(z,{\bar z})=0$.  Therefore we can separate
the holomorphic  and antiholomorphic parts, $\sigma(z,{\bar z})\equiv\sigma(z)+
{\bar \sigma}({\bar z})$.  The 2-point functions of $\sigma(z)$ and
${\bar \sigma}({\bar z})$  can be normalized as
\eqn\www{<\sigma(z)\sigma(w)>=2\log(z-w),~~~
<{\bar \sigma}({\bar z}){\bar \sigma}({\bar w})> = 2\log({\bar z}
-{\bar w}).}
Hence $h$ becomes
\eqn\eq{ h=\exp\left(-i\sigma(z)/{\sqrt{k}}\right)
\exp\left(-i{\bar \sigma}({\bar z})/{\sqrt{k}}~.\right)}
A chiral gauge transformation, $g_L^{-1}({\bar z})=\exp\left(-i{\bar
\sigma}({\bar z})/{\sqrt{k}}\right)$, can be used to remove the gauge degree
of freedom such that $h(z,{\bar z})$ becomes a holomorphic function,
$h_{\rm p}(z)$,
\eqn\C{
h_{\rm p}(z)=\exp\left(-i\sigma(z)/{\sqrt{k}}\right).}
The $U(1)$  current  $J_h$ from the action Eq.\free\  is
defined by (using the formula shown in the paragraph below Eq.\gltr\ )
\eqn\PPq{ J_h(z)={-1\over{2}}\cdot (-k)\cdot\partial_z h\cdot h^{-1}
={-i\over{2}}\sqrt{k}\partial_z\sigma(z) }
which satisfies $U(1)$ KM algebra at the level ``$-k$'', justifying the action
Eq.\free\ we started with,
\eqn\eq{
J_h(z)J_h(w)={-k/2\over{(z-w)^2}} + {\rm reg.~terms}.}
$J_h$ commutes with the original $SU(2)_k$ KM algebra, \ie\
\eqn\eq{ J_h(z)J_{\phi}^{3,\pm}(w)={\rm reg.~terms}.  }
The $\sigma$ field has the stress energy-momentum tensor $T_{\sigma}(z)=
{1\over 4}
\partial_z \sigma\partial_z \sigma$ with central charge $c_{\sigma}=1$.
The $b,c$ ghost action in Eq.\quan\ has the stress energy-momentum tensor
$T_{bc}(z)=-b\partial_z c$ whose central charge is $c_{\rm gh}=-2$.
Here $J_{\rm gh}$ vanishes because of the abelian nature of $H_L$.  Therefore,
\eqn\eq{\eqalign{  J_{\rm tot}^3(z)&= J_{\phi}^3(z)+J_h(z)\cr
{\rm where}~~~~J_{\rm tot}^3(z)&J_{\rm tot}^3(w)\sim {\rm reg.~terms}.\cr}}
Note that the KM anomaly term is absent as is required by the gauge invariance.
The stress energy-momentum tensor of the left-moving gauged WZW theory
 is obtained as
\eqn\H{ T_{\rm tot}(z)=T_{SU(2)}(z)+T_{\sigma}(z)+T_{bc}(z) }
whose central charge is $c_{\rm tot}=3k/(k+2)+1-2=2(k-1)/(k+2)$.
The above discussion applies equally to the right-handed sector if
$H_R=U(1)$ also, where
\eqn\B{
{\widetilde h}_{\rm p}({\bar z})=\exp\left(-i{\widetilde \sigma}({\bar z})/
\sqrt{k}\right)~~~{\rm and}~~~J_{\tilde h}({\bar z})={-i\over{2}}\sqrt{k}
\partial_{\bar z} {\widetilde \sigma}({\bar z})~. }

Following from the discussion between Eq.\A\ and \JANseven , we have to
solve
\eqn\F{\eqalign{
\left(J_{\phi}^{3}(z) +J_{h}(z)\right)\Phi^j_m(w)&={\rm reg.~terms}\cr
\left({\bar J}_{\phi}^{3}({\bar z}) +J_{\tilde h}({\bar z})\right)
{\bar \Phi}^{\bar j}_{\bar m}({\bar w})&={\rm reg.~terms}\cr }}
where
\eqn\FF{
\Phi^j_m(z)=F_1(h_{\rm p})(z)\cdot G^j_m(z)~,~~~~{\bar \Phi}^{\bar j}_{\bar m}
({\bar z})=F_2({\widetilde h}_{\rm p})({\bar z}){\bar G}^{\bar j}_{\bar m}
({\bar z})~.}
The answers are
\eqn\eq{\eqalign{
\Phi^{j}_{m}(z) &= G^j_m(z)\exp\left(-im\sigma(z)/\sqrt{k}\right)= G^j_m(z)
{h_{\rm p}^m}(z)\cr
{\bar \Phi}^{\bar j}_{\bar m}(z) &= {\bar G}^{\bar j}_{\bar m}({\bar z})
\exp\left(-im{\widetilde \sigma}({\bar z})/\sqrt{k}\right)
={\bar G}^{\bar j}_{\bar m}({\bar z}){{\widetilde h}_{\rm p}^m}({\bar z})~.
\cr}}
The holomorphic part of the correlation functions of these dressed primary
fields can be calculated as
\eqn\eq{\eqalign{
&<\Phi^{j_1}_{m_1}(z_1)\Phi^{j_2}_{m_2}(z_2)...\Phi^{j_n}_{m_n}(z_n)>
=<G^{j_1}_{m_1}(z_1)G^{j_2}_{m_2}(z_2)...G^{j_n}_{m_n}(z_n)>\cdot \cr
   &~~~~<e^{\left({-im_1\sigma(z_1)/{\sqrt{k}}}\right)}
e^{\left({-im_2\sigma(z_2)/{\sqrt{k}}}\right)}...
e^{\left({-im_n\sigma(z_n)/{\sqrt{k}}}\right)} >\cr }}
where the first factor is the correlation function of $SU(2)_k$ WZW theory.
%The antiholomorphic parts can be similarly calculated.
  These correlation functions are exactly the same
as those in the $Z_k$ parafermion (PF) theory\ZFref .  To show this more
explicitly,
we can use the $Z_k$ PF as a particular parametrization of the $SU(2)_k$
KM currents (the antiholomorphic part is again omitted),
\eqn\eq{ \eqalign{
J^+(z)&=\sqrt{k}\psi_1(z)\exp\left(i\rho(z)/\sqrt{k}\right)\cr
J^-(z)&=\sqrt{k}\psi_1^{\dagger}(z)\exp\left(-i\rho(z)/\sqrt{k}\right)\cr
J^3(z)&={i\over{2}}\sqrt{k}\partial_z\rho(z)\cr }}
where $\psi_1$ and $\psi_1^{\dagger}$ are the PF currents which are the
symmetry currents in the $Z_k$ PF theory.  $\rho(z)$ is a free space-like
boson with
the two point function
\eqn\wwww{ <\rho(z)\rho(w)>=-2\log(z-w)~.}
The central charge of this $Z_k$ PF is $c_{\rm PF}=2(k-1)/(k+2)$ which
is the same as that for Eq.\H .
The primary fields ${\Psi^j_m}$ in the $Z_k$ PF theory are also related to
those in the WZW theory as
\eqn\eq{
G^j_m(z)={\Psi}^j_m(z) \exp\left(im\rho(z)/\sqrt{k}\right) }
whose highest weight is
\eqn\G{{\Delta}\left({\Psi}^j_m\right)=\Delta^j-m^2/k~.}
Hence it can be shown that
\eqn\eq{\eqalign{
&<\Phi^{j_1}_{m_1}(z_1)\Phi^{j_2}_{m_2}(z_2)...\Phi^{j_n}_{m_n}(z_n)>\cr
&=<{\Psi}^{j_1}_{m_1}(z_1){\Psi}^{j_2}_{m_2}(z_2)...
{\Psi}^{j_n}_{m_n}(z_n)>
<e^{\left({im_1\rho(z_1)/{\sqrt{k}}}\right)}
e^{\left({im_2\rho(z_2)/{\sqrt{k}}}\right)}...
e^{\left({im_n\rho(z_n)/{\sqrt{k}}}\right)}>\cr
 &~~~~<e^{\left({-im_1\sigma(z_1)/{\sqrt{k}}}\right)}
e^{\left({-im_2\sigma(z_2)/{\sqrt{k}}}\right)}...
e^{\left({-im_n\sigma(z_n)/{\sqrt{k}}}\right)}>\cr }}
where
\eqn\eq{<e^{\left({-im_1\sigma(z_1)/{\sqrt{k}}}\right)}
e^{\left({-im_2\sigma(z_2)/{\sqrt{k}}}\right)}...
e^{\left({-im_n\sigma(z_n)/{\sqrt{k}}}\right)}>=
\prod^n_{i=1}\prod_{j>i} (z_i-z_j)^{-2m_im_j/k} }
and
\eqn\eq{ <e^{\left({im_1\rho(z_1)/{\sqrt{k}}}\right)}
e^{\left({im_2\rho(z_2)/{\sqrt{k}}}\right)}...
e^{\left({im_n\rho(z_n)/{\sqrt{k}}}\right)}>=
\prod^n_{i=1}\prod_{j>i} (z_i-z_j)^{2m_im_j/k} }
according to the 2-point functions in Eq.\www\ and \wwww .
This demonstrates that
\eqn\eq{
<\Phi^{j_1}_{m_1}(z_1)\Phi^{j_2}_{m_2}(z_2)...\Phi^{j_n}_{m_n}(z_n)>
=<{\Psi}^{j_1}_{m_1}(z_1){\Psi}^{j_2}_{m_2}(z_2)...
{\Psi}^{j_n}_{m_n}(z_n)>~.}
Note that the highest weight of $\Phi^j_m(z)$ can be calculated as
\eqn\eq{
\left(T_{SU(2)}(z)+T_{\sigma}(z)+T_{\rm gh}(z)\right)\Phi_m^j(w)
={(\Delta_j-m^2/k)\over{(z-w)^2}}\Phi_m^j(w)+... }
which is the same as that in Eq.\G .
This establishes the isomorphism between the chiral gauged WZW theory
and  the $Z_k$ PF theory in the CFT.  In this picture of the
gauged WZW theory,  the gauge
fields form a ``cloud'' surrounding the primary fields in the WZW theory
and screening the $U(1)$ subgroup  of the $SU(2)$ WZW theory.
 According to Eq.\www , $h_{\rm p}$ is treated as a non-unitary theory
based on a time-like boson, in contrast to the space-like boson defined
in  Eq.\wwww .  In other words, the gauge fields cancel
one physical unitary degree of freedom from the original WZW theory, leaving
the unitary $Z_k$ PF behind.
The original $SU(2)$ symmetry is no longer present in the gauged WZW theory,
which can be understood when applying $J^{\pm}_0$ to $\Phi_m^j$.
The resulting states, $J^{\pm}_0\Phi^j_m(z)=G^j_{m\pm 1} h_{\rm p}^m$,
 do not satisfy the physical state condition, Eq.\F .
This implies that the  ``physical currents'',
$J_{\rm p}^{\pm}$, in the chiral gauged WZW theory are also screened by
a cloud of gauge fields.  The simple requirement is
\eqn\eq{ J_{\rm p}^{\pm}(z) \Phi^j_m(w) \sim \Phi^j_{m\pm 1}(w)~.}
This implies
\eqn\eq{ J_{\rm p}^{\pm}(z)= J^{\pm}(z)\exp\left(\mp i\sigma (z)/\sqrt{k}
\right). }
It can be checked that
\eqn\eq{ J^3_{\rm tot}(z) J_{\rm p}^{\pm}(w) = {\rm reg.~ terms} }
which means that the negative modings of $J_{\rm p}^{\pm}$ can be used to
act on the physical primary fields and  generate the descendants in the
gauged WZW theory.
The OPE between $J_{\rm p}^{+}(z)$ and $J_{\rm p}^{-}(w)$ can be calculated as
\eqn\eq{\eqalign{
J_{\rm p}^{+}(z)J_{\rm p}^{-}(w)&=\left(J^+(z)J^-(w)\right)
\left(\exp\left(-i\sigma(z)/\sqrt{k}\right)
      \exp\left(i\sigma(w)/\sqrt{k}\right)  \right)\cr
&=\left({k\over{(z-w)^2}}+{2J^3(w)\over{z-w}}+...\right)
(z-w)^{2/k}\left[1-{i\over{\sqrt{k}}}(z-w)\partial\sigma(w)\right]\cr
&=(z-w)^{-2+2/k}\left[1+...\right]\cr }}
which is isomorphic to  that of the $Z_k$ PF currents $\psi_1$ and
$\psi_1^{\dagger}$.
This concludes our argument that the chiral $U(1)$ gauged $SU(2)_k$ WZW theory
is isomorphic to the $Z_k$ PF theory.  The above argument can easily generalize
to the case where both $H_L$ and $H_R$ are abelian.

\medskip
{\noindent{\bf (II.)~${\bf G}=SU(2)_k\otimes SU(2)_l,~{\bf H}=SU(2)$:}}
\smallskip

In the case where $G$ is not simple but semi-simple, the above analysis
must be slightly modified.  To be specific, let us consider the case when
$G=SU(2)_k\otimes SU(2)_l$ where both $k$ and $l$ are integers.
In this case, the WZW action can be written as
\eqn\eq{ S(\phi_1, \phi_2)=kI(\phi_1)+lI(\phi_2).}
The chiral gauge coupling, $H=SU(2)$, can be introduced as that in Eq.\ftwo\
\eqn\eq{ S(A^L_{\bar z},A^R_z,\phi_1,\phi_2)=S(A^L_{\bar z},
A^R_z,\phi_1)+ S(A^L_{\bar z},A^R_z,\phi_2)~. }
Then Eq.\chang\ for this case becomes
\eqn\JJAN{ S(h,{\widetilde h},\phi_1,\phi_2)=kI({\widehat \phi}_1)+l
I({\widehat \phi}_2)-(k+l)I(h)-(k+l)I({\widetilde h}).}
The \KM\ currents from $\phi_1,\phi_2,h_{\rm p}$ and ${\widetilde h}_{\rm p}$
and the  associated Sugawara stress energy-momentum tensors can be similarly
constructed as discussed before.  Therefore, we can first calculate the
total central charge for the holomorphic sector,
\eqn\eq{\eqalign{ c_{\rm tot}&={3k\over{k+2}}+{3l\over{l+2}}+
{3(-k-l-4)\over{-k-l-2}}-2\times 3\cr
&={3k\over{k+2}}+{3l\over{l+2}}-{3(k+l)\over{k+l+2}}\cr}}
which is the same as that of the $\SU $ theory.
Before the introduction of gauge couplings, the primary field of $SU(2)_k\times
SU(2)_l$ is
\eqn\eq{
G^{j_1j_2{\bar j}_1{\bar j}_2}(z,{\bar z})
=\left(G^{j_1}(z)G^{j_2}(z)\right)
\left({\bar G}^{{\bar j}_1}({\bar z})G^{{\bar j}_2}({\bar z}) \right) }
where the $m$ quantum number is suppressed, \ie\ $G^{j}(z)=G^{j}_j(z)$.

Let us consider the sector $\left[ G^{j_1}(z)G^{j_2}(z) \right]$ which
includes both the primary field and its current algebra descendants.
{}From Eq.\JJAN , we see that the gauge field's contribution behaves like a
$SU(2)_{k+l}$ WZW theory with time-like metric.  Since the
$\left[ G^{j_1}(z)G^{j_2}(z) \right]$ sector is reducible in terms of the
representations of $SU(2)_{k+l}$ WZW theory, this means the
$\left[ G^{j_1}(z)G^{j_2}(z) \right]$ sector contains  subsectors, each
of which will have different gauge dressing  when we turn on the gauge
couplings.  Therefore, we have to decompose the sector
$\left[ G^{j_1}(z)G^{j_2}(z) \right]$ into subsectors
\eqn\eq{\left[ G^{j_1}(z)G^{j_2}(z)\right]=\sum_j V_{j_1j_2}^j(z) }
where generically $j=0,1/2,...,(k+l)/2$, with the restriction
$j=j_1+j_2~(mod~1)$.
Therefore the conformal dimension of $V_{j_1j_2}^{j}(z)$ is
\eqn\eq{\Delta \left( V_{j_1j_2}^{j}\right)={j_1(j_1+1)\over{k+2}}
  + {j_2(j_2+1)\over{l+2}}+N^j_{j_1j_2} }
where $N^j_{j_1j_2}$ is a non-negative integer.

 {} Following the discussion between Eq.\A\ and \JANnine , we find the
dressed primary fields to be
\eqn\eq{ \Phi^j_{j_1j_2}(z)= V^j_{j_1j_2}(z) H^j_{j_1+j_2}(z) }
where the dressing field $H^j_{j_1+j_2}(z)$ is a time-like primary field
in the spin $j$ representation of $SU(2)_{k+l}$ theory, and
 $H^j_{j_1+j_2}(z)$ has the conformal dimension
$-j(j+1)/(k+l+2)$.  Hence the conformal dimension of
$\Phi^j_{j_1j_2}(z)$ is given by
\eqn\eq{ \Delta(\Phi^j_{j_1j_2})={j_1(j_1+1)\over{k+2}}+{j_2(j_2+1)\over{l+2}}
-{j(j+1)\over{k+l+2}} +N^j_{j_1j_2} }
which is the same as that of the primary field in the $SU(2)$ coset theory.
The holomorphic correlation function of these dressed
primary fields are given as
\eqn\DIEtwo{\eqalign{
<\Phi&^{j(1)}_{j_1(1)j_2(1)}(z_1)\Phi^{j(2)}_{j_1(2)j_2(2)}(z_2)...
\Phi^{j(n)}_{j_1(n)j_2(n)}(z_n)>=\cr
&<V^{j(1)}_{j_1(1)j_2(1)}(z_1)V^{j(2)}_{j_1(2)j_2(2)}(z_2)...
V^{j(n)}_{j_1(n)j_2(n)}(z_n)>\cdot\cr
&<H^{j(1)}_{j_1(1)+j_2(1)}(z_1)H^{j(2)}_{j_1(2)+j_2(2)}(z_2)...
H^{j(n)}_{j_1(n)+j_2(n)}(z_n)>\cr } }
which is the same as that of the $\SU$ coset theory.
To see that
$<\Phi^{j(1)}_{j_1(1)j_2(1)}(z_1)\Phi^{j(2)}_{j_1(2)j_2(2)}(z_2)
...\Phi^{j(n)}_{j_1(n)j_2(n)}(z_n)>$ is the correlation function for the coset
theory, we can decompose the WZW primary field,
\eqn\eq{ V^{j}_{j_1j_2}(z)=g^j_{j_1j_2}(z)G^j_{j_1+j_2}(z)}
where $g^j_{j_1j_2}(z)$ is the coset primary field and
$G^j_{j_1+j_2}(z)$ is the primary field with $m$ quantum number $j_1+j_2$
in the spin-j representation of
$SU(2)_{k+l}$ WZW theory, whose conformal dimension is $j(j+1)/(k+l+2)$,
which is exactly opposite to that of $H^j_{j_1+j_2}(z)$.
{} From the  argument leading to Eq.\JANtwelve , we can see that the
holomorphic correlation function, Eq.\DIEtwo , is precisely that of
the $g^{j(i)}_{j_1(i)j_2(i)}(z_i)$, \ie\ the holomorphic correlation
function of primary fields of the $\SU$ coset theory.

\newsec{\bf Explicit Representations of the Matter Fields}

In this section, two explicit examples are briefly discussed to illustrate
some of the issues in the chiral gauged theory.
The first one is the $SO(N)$ representation of real fermions in which we can
see why chiral
gauge coupling is anomaly-free. The other one is a free
spin $(1,0)$  system of ${\omega^{\dagger}}^a$, $\omega^a$ in the adjoint
representation of $G_L$ and couples to chiral gauge fields in
$H_L=G_L$; this is an example of the critical $k$ case
in which the $h_{\rm p}$ fields decouple.

As an example of Eq.\Faction , real fermions with the chiral gauge
couplings in two dimensions can be
introduced with the action
\eqn\DEC{
S={1\over{8\pi}}\int  \psi^i(\delta^{ij}\partial_{\bar z}+
{A^L_{\bar z}}^a t^a_{ij})\psi^j }
where $i=1,2,...,N$ for $N\geq 4$.
We can consider these fermions forming a vector representation of $SO(N)$
with the representation matrix $t^a$.  $a=1,2,...,dim\left(SO(N)\right)$
for each generator.  The affine KM currents can be constructed as
\eqn\eq{ J^a(z)\equiv \psi^i(z) t^a_{ij} \psi^j (z) }
satisfying $SO(N)$ affine KM algebra at the level $k=1$.  Here the gauge
couplings are chiral, \ie\ in Eq.\DEC\ $A^L=\left( 0, A^L_{\bar z}\right)$
and $A^R$ is absent.

Before we consider the vacuum polarization of this theory,
let us first consider the general current-current correlation
function, $<J_{\mu}^a J_{\nu}^b>=\Pi^{ab}_{\mu\nu}=\delta^{ab}\Pi_{\mu\nu}$,
 which has the  general form,
\eqn\Dec{{\Pi}_{\mu\nu}(q)=g_{\mu\nu}\Pi_1(q)+{q_{\mu}q_{\nu}
\over{q^2}}
\Pi_2(q)+{\epsilon_{\mu\lambda}q^{\lambda}q_{\nu}+\epsilon_{\nu\lambda}
q^{\lambda}q_{\mu}\over{2q^2}}\Pi_3(q) }
where $q_{\mu}$ is the momentum and $\Pi_i(q)$ are
functions of $q^2$.  As is clear from the form of
Eq.\Dec , it is impossible
for all (vector, axial, chiral) currents to be conserved.
For example, the requirement of the vector current conservation demands
\eqn\ONE{ q^{\mu}\Pi_{\mu\nu}(q)=0 \Rightarrow \Pi_1(q)=-\Pi_2(q)~~~~{\rm and}
   ~~~~ \Pi_3(q)=0 }
while the conservation of the axial-vector current implies $q_{\lambda}
\epsilon^{\lambda\mu}\Pi_{\mu\nu}(q)=0$ which demands
\eqn\TWO{ \Pi_1 = 0 ~~~~{\rm  and}~~~~ \Pi_3 = 0~.}
Hence the demand of both conservation implies $ \Pi_1 = \Pi_2 = \Pi_3 =0$.
An examination of the one-loop diagram shows that we can regularize the
one-loop effect such that either \ONE\ or \TWO\ can be satisfied.
This is sometimes called the renormalization condition.  But there does
not exist a regularization choice such that all $\Pi_i(q)$ vanishes.
Therefore if the fermions are coupled only to vector or only to axial vector
currents, the gauge invariance can be maintained, but not when coupled to
both.

For the theory Eq.\DEC , only
$J^a_z$ couples to $A^L_{\bar z}$, so ${\Pi}_{zz}^{ab}$ is the vacuum
polarization from the fermion loop contributions.
The chiral gauge invariance requires that
\eqn\eq{ 0= q_{\bar z} \Pi_{zz}(q)={q_z\over{2}}\left[\Pi_2(q)-\Pi_3(q)
\right] }
where the coordinates $g^{z{\bar z}}=1$ and $\epsilon^{z{\bar z}}=1$ are
used.  Therefore, the renormalization condition can be chosen as
$\Pi_2(q)=\Pi_3(q)$ such that the chiral gauge invariance is preserved.
Since  in this case  $C_2(SO(N))=N-2$, the full quantum action has to include
$h_{\rm p}$ field as discussed in Sec.3.

The next example is the free ghost system of spins $(1,0)$,
${\omega^{\dagger}}^a$, $\omega^a$.  Their action is
\eqn\SLaction{ S_L=\int d^2z\prod^{dim G_L}_{a=1}
          ~{\omega^{\dagger}}^a~ {\bar \partial}~\omega^a~.}
The equations of motion are ${\bar \partial}\omega^a=0
={\bar \partial}{\omega^{\dagger}}^a$. The non-trivial OPEs between them are
\eqn\OPEex{ \omega^a(z)~ {\do}^b(w) = {\delta^{ab}\over{z-w}}
                             =-~{\do}^a(z)~\o^b(w)}
and the rest are regular.  The action, \SLaction , has a local symmetry
determined by
\eqn\globeq{ \eqalign{& g_L(z)\equiv \exp\left(\vraised_L(z)\right)~~~
        {\rm with}~~~\vraised_L(z)_{bc} =\epsilon^a(z)f^{a}_{bc} \cr
&{\rm and~~~~~~~~~~~~} {\omega^{\dagger}}\rightarrow {\omega^{\dagger}}
    g_L^{-1}(z)~,~~~~{\omega}\rightarrow g_L(z){\omega}~.\cr } }
The infinitesimal variations of the fields are
\eqn\eq{ \eqalign{ \delta^a \o^b(z) &= \e^a(z) f^a_{bc} {\o}^c(z) \equiv
\e^a(z)\left[Q^a~,~~\o^b(z)\right]\cr
  \delta^a {\do}^b(z) &= \e^a(z) f^a_{bc} {\do}^c(z)\equiv
            \e^a(z)\left[Q^a~,~~{\do}^b(z)\right]\cr }}
where $Q^a$ is the generator of the global transformation in Eq.\globeq\
and the index $a$ is not summed over.  From the OPEs in Eq.\OPEex ,
we can easily solve $Q^a$ as the contour integral of the affine
KM symmetry current
\eqn\eq{ Q^a \equiv \oint dz J^a =\oint dz f^{abc}\o^b {\do}^c}
These currents satisfy
\eqn\eq{ J^a(z) J^b(w) = {-C_2(G_L)\over{(z-w)^2}} + {f^{abc}
J^c(w)\over{z-w}}+....}
which is the affine KM algebra at level $k=-2C_2(G_L)$.
We can now introduce the chiral gauge symmetry, \ie\ replacing $\e^a(z)$ by
$\e^a({\bar z})$.
The covariant derivative can be introduced as $D_{\bar z}\equiv
\partial_{\bar z}+ A^L_{\bar z}$.  The action becomes
\eqn\eq{S_L=\int d^2z\prod^{dim G_L}_{a=1}
          ~{\omega^{\dagger}}^a~ \left({\bar \partial}+A^L_{\bar z}\right)
{}            ~\omega^a~.}
The invariance of the action
requires  the gauge transformation on $A^L_{\bar z}$ as
\eqn\eq{ A^L_{\bar z} = g_L~A^L_{\bar z}~ g_L^{-1}-\partial_{\bar z}g_L
             ~ g_L^{-1}~.}
Since $k=-2C_2(G_L)$, which  is the critical value that $h_{\rm p}$ decouples
from the matter system.  Therefore the gauge fixed partition function
with Fadeev-Popov ghosts ($b^a,~c^a$) is
\eqn\OME{\eqalign{ {\cal Z}_{\rm gf}&=\int {\cal D}{\do}^a{\cal D}{\o}^a
{\cal D}{b}^a {\cal D}{c}^a \exp\left(-S_{gf}\right)\cr
{\rm where}&~~~S_{\rm gf}=\int d^2z \left({\do}^a{\bar \partial}\o^a
     + b^a{\bar \partial}c^a\right)~.}}
The full quantum action has the BRST symmetry which can be obtained from
Eq.\globeq\ with the well-known modification, $\e^a \rightarrow \lambda c^a$,
($\lambda$ is a constant grassman number.)
\eqn\eq{ \eqalign{
\delta~\o^b &= \lambda c^a \delta^a \o^b =  \lambda c^a f^{abc}\o^c
          \equiv [Q_{BRST}~,~\o^b]\cr
\delta~{\do}^b &=\lambda c^a \delta^a {\do}^b=\lambda c^a f^{abc}{\do}^c
          \equiv [Q_{BRST}~,~{\do}^b]\cr
\delta~c^{\b} &= {1\over 2} \lambda c^a f^{a\b c}\c^c
           \equiv \{ Q_{BRST}~,~c^{\b} \} \cr
\delta~b^{\b} &= \lambda f^{\b\a\c} \o^{\a} {\do}^{\c}
         +\lambda c^{\a} f^{\a\b\c}b^{\c} \equiv \{Q_{BRST}~,~b^{\b} \}~.\cr}}
where  $\o , \do , b, c$'s dependence in $z$ is suppressed.
It is an easy check that $\delta S_{\rm gf}=0$.
The BRST current and operator can be found if we demand the above variations
as the results of the (anti-)commutation relations between the BRST oerators
and various fields.  The simple calculation from the OPE in \OPEex\ gives
\eqn\eq{ \eqalign{
Q_{BRST}&=\oint J_{BRST}(z)\cr
J_{BRST}(z)&=\sum_{\a\b\c}c^{\a}f^{\a\b\c}\o^{\b}{\do}^{\c}(z) -
{1\over 2}f^{\a\b\c} c^{\a}c^{\b}b^{\c}(z)\cr }}
This is exactly the same BRST operator we found from the purely abstract
gauged WZW theory, \Zvon .  Note that this model has $G_L=H_L$
while $G_R$ and $H_R$ are absent.

It is natural to ask whether this model is unitary or not.  The $\o -\do$
theory without gauge fields is of course non-unitary.  However, of all the
states in the gauged $\o -\do$ theory, Eq.\OME , only a small subset of
them satisfy the physical state conditions (or equivalently, BRST
cohomology).  This situation is very similar to the string
theory case.  Recall that the conformal theory of $26$ bosons with Minkowski
metric is non-unitary.  However, in the bosonic string theory, the spectrum
must satisfy the physical state conditions.  As a result, the string theory,
which is the gauged (with the metric) version of the $26$-boson conformal
theory, is unitary.  Following from this analogy, it is likely that
the gauged $\o -\do$ theory is
also unitary.  A preliminary analysis seems to support this conjecture\ACT .

\newsec{\bf Conclusion}

In this paper, we showed that the chiral gauged WZW theory is anomaly free and
its quantization can be carried out. In the special case where the left
and the right chiral gauge groups are  the same, the theory is in agreement
with the vector gauged WZW theory.  However, the chiral gauged WZW theory
has the advantage of allowing the existence of  the left-right asymmetrical
gauge groups which may be suitable for the heterotic type model building in
the fractional superstring theory.  By an examination of correlation
functions, we argue that the  gauged WZW theory in the conformal
limit is exactly the coset theory in CFT.

The examples given in Sec.5,  can be rewritten as gauged WZW theories,
in which case $G=G_L=G_R$.  In the first example of real fermions,
the non-linear sigma field $\phi$ is in the vector representation
of $G_L$ and the identity representation of $G_R$; in the second
example, $\phi$ is in the adjoint representation of
$G_L$ and the identity representation of $G_R$.  In both cases, the left gauge
group is $H_L=G_L$ while $H_R$ is  absent.  From this point of view, the
chiral gauged WZW theory is more general than the vector gauged WZW theory.

In superstring theory, the Fadeev-Popov ghost system (and the related BRST
operator) is obtained by gauge fixing the two-dimensional supergravity
action which has a world-sheet local supersymmetry.  To understand the
ghost system
in the fractional superstring case, it will be nice to find out the world-sheet
local fractional supersymmetry from a two-dimensional action which involves
the parafermion.  This requires, as a first step,
 a classical action for the parafermion,
which can be provided by the chiral gauged WZW theory.  It is very
similar to anyons in $2+1$ dimensions where a classical action for anyons
requires the introduction of the Chern-Simon gauge coupling into the action
for particles with integer spin or half-integer spin.

Finally the analogy between the gauged WZW
theory and Polyakov's bosonic string theory as summarized in Table 2 is
striking.

\bigskip
\medskip

\moveleft8pt
\vbox{\halign{\hfil{\footnotefont\bf#}&\quad\hfil{\footnotefont\it#}\hfil&
             \quad\hfil {\footnotefont\it#}\hfil\cr
      {\phantom{symmetry}}&{\bf Gauged WZW Theory}&
                                     {\bf Polyakov's String Theory}\cr
       Gauge Field&Gauge connections&Two dimensional metric\cr
       Gauge Symmetry&{$\vraised_L({\bar z})$ and $\vraised_R(z)$}&
                      Diffeomorphism $\xi^z~,\xi^{\bar z}$\cr
    Ghost System&{$c^a,b^a_z, a=1,2,...,dim H_L$}&{$c^z, b_{zz}$}\cr
    {\phantom{symmetry}}&{${\bar c}^{\a},{\bar b}^{\a}_{\bar z}, \a =1,2,...
       dim~H_R$}&   {$c^{\bar z}, b_{{\bar z}{\bar z}}$}\cr
    {\phantom{symmetry}}& {$dim(c)=0,~ dim(b)=1$}&{$dim(c)=-1,~ dim(b)=2$}\cr
    Remaining Fields& {$h_{\rm p}$ and ${\widetilde h}_{\rm p}$}&
   {Liouville field,~$\phi (z,{\bar z})=\phi (z)+{\bar \phi}({\bar z})$}\cr
     BRST Operator&{$Q^L_{BRST}=\oint c^a\left( J^a+J^a_h+J^a_{gh}\right)$}&
                     {$Q_{BRST}=\oint c\left( T_m+T_{\phi}+T_{gh}\right)$}\cr
{\phantom{symmetry}}&{$Q^R_{BRST}=\oint {\bar c}^{\a}\left( {\bar J}^{\a}
              +{\bar J}^{\a}_{\tilde h}+{\bar J}^{\a}_{gh}\right)$}&
  {${\bar Q}_{BRST}=\oint {\bar c}\left( {\bar T}_m+{\bar T}_{\bar \phi}
                          +{\bar T}_{gh}\right)$}\cr
Critical Value&{$k=-2C_2(H_L)=-2C_2(H_R)$}&{$c_m={\bar c}_m=26$}\cr
Matter Fields&{$G^{\Lambda{\bar \Lambda}}_{\lambda{\bar \lambda}}$}&
{$\Phi_{mn}$}\cr
Dressed Fields&{$\Phi^{\Lambda{\bar \Lambda}}_{\lambda{\bar \lambda}}
    =G^{\Lambda{\bar \Lambda}}_{\lambda{\bar \lambda}} F(h_{\rm p},~
   {\widetilde h}_{\rm p})$}&{$\widetilde \Phi_{mn}=\Phi_{mn}F(\phi)$}\cr }}
\bigskip
\centerline{
\vbox{
\hsize=3.0in
\item{\bf Table 2.} Similarities between the gauged
WZW theory and Polyakov's bosonic string theory. }}

\bigskip

In Table 2, $\Phi_{mn}$ are primary fields in the unitary minimal models
 with
\eqn\eq{ c=1-{6\over{p(p+1)}}. }
See Ref.\ddk\ for details.

\centerline{\bf Acknowledgements}

It is a pleasure to thank P.C. Argyres, D. Karabali, A. LeClair and
H.T. Schnitzer for valuable discussions.
This work was supported in part by the National Science Foundation.

\appendix{A}{}
 In this Appendix we review the propreties of the non-abelian anomaly in
$2n$-dimensions, using the  differential geometry approach.
 The chiral anomaly was first derived in Ref.\Zumino\ while the relation
between the anomaly in LR-scheme and that in the A-scheme was first
clarified in Ref.\KT . For a more comprehensive
review, see Ref.\refs{\Petersen,\AG}.  Our
purpose of including this appendix is to clarify the notations and  make
the  paper self-contained.  This review is longer than needed to derive the
non-abelian anomaly, because the formalism will be needed in Appendix B.

The expressions of the
anomalies involve only  gauge fields and gauge transformation parameters
since matter fields have been integrated out.
The effective action in $2n$-dimensions, $W(\Apair)$,  can be written as an
integral in $2n+1$-dimensions\WZref\
\eqn\anomalythree{ W (\Apair)=c_n \int_{B^{2n+1}} \anom (\Apair)~,}
where $\anom$ is a $(2n+1)$-form.  The superscript indicates the power of
$\vraised_L$ and $\vraised_R$.  So $\anom$ does not depend on
$\vraised_L , \vraised_R$ explicitly.
We are interested in the space,
$S^{2n}$, which is the boundary of $B^{2n+1}$.
The consistency between Eq.\anomaly\ and
\anomalythree\
requires that
\eqn\crucial{\df \omega ^1_{2n}(\Apair;\vpair)= \var\anom(\Apair)~.}
The relation between the non-abelian anomaly in $2n$-dimensions and the
abelian anomaly $\Omega_{2n+2}(\Apair)$ in $(2n+2)$-dimensions was first
shown in Ref.\Zumino :
\eqn\manone{\eqalign{
\Omega_{2n+2}(\Apair)&=\df \anom (\Apair) \cr
\Omega_{2n+2}(\Apair)={\rm Tr} \Bigl( F_R^{n+1}&-F_L^{n+1} \Bigr)
=\int^{t_2}_{t_1}dt {d\over{dt}} {\rm Tr} \left[ F(t)^{n+1}\right]\cr } }
with the boundary condition specified by
\eqn\eq{{
A(t_1)=A^L~,~~A(t_2)=A^R~~~~~{\rm and}~~~F(t)\equiv \df A(t)+A(t)^2~.}}
(Again, all products are wedge products.)
Exploiting properties of the differential form we
can rewrite the second equation of \manone\ as
\eqn\man{\eqalign{\Omega_{2n+2}(\Apair)&=\int^{t_2}_{t_1} dt (n+1){\rm Tr}
\left( {\dot F}(t) F^n(t) \right) \cr
&=(n+1)\int^{t_2}_{t_1} dt~ {\rm Tr}\left\{\df{\dot A}(t)F^n(t)
+{\dot A}(t)[A(t),F^n(t)]\right\}\cr
&=(n+1)\int^{t_2}_{t_1} dt ~\df~ {\rm Tr}\left\{ {\dot A}(t)F^n(t)\right\}~, }}
where we have used
$$\vbox{\halign{\indent#\hfil\cr
(i) {\rm Tr}~$(AB)$~=~$-$~{\rm Tr}~$(BA)$~~~~~{\rm if} $A$ {\rm and} $B$
{\rm are both odd forms~;}\cr
(ii) {\rm Bianchi~ identity, i.e.}~~ $DF~=~0~=~\df F+[A,F]$~.\cr}}$$
Comparing Eq.\manone\ with  \man\ we get
\eqn\route{\anom (\Apair , \gamma)=(n+1)\int^{t_2}_{t_1} dt~{\rm Tr}~\left\{
{\dot A}(t) F^n(t) \right\}~.}
Here a possible path choice, $\gamma$, in the expression of
$\anom$ is introduced because  all previous discussions
up to Eq.\route\ depend only on the end points of the $t$-integration.
Different choices of
paths can be related to different choices of renormalization conditions.

There are two renormalization conditions frequently used.  One is called
the A-scheme in which the vector
gauge transformation is invariant while the axial
vector gauge transformation is anomalous.  The other is
called the LR-scheme in which the anomaly, $\omega_{2n}^1$, and the effect
action, $W$, can be separated into the left and  the right pieces
independently.  In the following we will specify the paths $\gamma_A$
(for the A-scheme) and $\gamma_{LR}$ (for the LR-scheme).  Hence the explicit
expressions of  $\omega_{2n+1}^0(\Apair)$ and $\omega^1_{2n}$ can be evaluated
by using Eq.\route\ and \crucial .
First of all, $\gamma_A$ is chosen as in Fig.1, which is parametrized by
%%%%%%%%%%%%%%%%%%%%%%%%%  FIG One  %%%%%%%%%%%%%%%%%%%%%%%%%%%%
$$\vbox{
\beginpicture
\setcoordinatesystem units <1pt,1pt> point at 0 0
\linethickness 8pt
\arrow <20pt> [.2,.4] from 100 0 to 180 0
\setlinear \plot 180 0 260 0 /
\put {$A^L$} [rb] <-1pt,10pt> at 100 0
\put {$\bullet$} at 100 0
\put {$\bullet$} at 260 0
\put {$A^R$} [lb] <1pt,10pt> at 260 0
\put {\twelvemi$\gamma_A$} <0pt,20pt> at 180 0
\put {{ \vbox{\hsize=310pt  \figufont \item{\bf Fig. 1.}
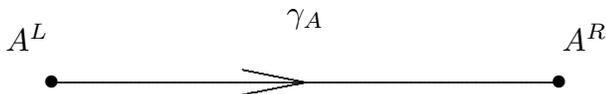
      The path $\gamma_A$ is defined by the parametrization
     $ A(t) \equiv V+tA, t\in [-1,1]$ where $V\equiv (A^L+A^R)/2$
 and $A\equiv (A^R-A^L)/2$.}   }} [lt] <-2pt,-3pt> at 10 -20
\endpicture          }$$
\medskip
%%%%%%%%%%%%%%%%%%%%%%%%%%%%%%%%%%%%%%%%%%%%%%%%%%%%%%%%%%%%%%%%
 $A(t) \equiv V+tA$, $t\in [-1,1]$  where $V\equiv (A^L+A^R)/2$
and $A\equiv (A^R-A^L)/2$.  With  this path, $\gamma_A$, we can easily obtain
\eqn\Ascheme{\anom(\Apair , \gamma_A)=(n+1)\int^1_{-1} dt~{\rm Tr}~[{1\over 2}
(A^R-A^L)F^n(t)]~.}  The vector gauge transformation is defined by
$v\equiv v_L\equiv v_R$.  And it is not hard to check the vector gauge
invariance, \ie ,
\eqn\eq{\delta_{v,v}\anom(\Apair , \gamma_A)=0~.}
In two dimensions, Eq.\Ascheme\ is equal to
\eqn\eq{ \omega_3^0(\Apair ,\gamma_{A})={\rm Tr}~\left((A^R-A^L)
\left[\df A^L+\df A^R\right]+{2\over 3}({(A^R)}^3-{(A^L)}^3) \right)~. }
Then from Eq.\crucial\ we can obtain
\eqn\eq{ \omega^1_{2,A}={\rm Tr}~\Bigl( (\df v_R-\df v_L)(A^R+A^L)
    +(v_R-v_L)(A^RA^L+A^LA^R) \Bigr)~. }

On the other hand, the path $\gamma_{LR}$ can be defined as in Fig.2,
\medskip
%%%%%%%%%%%%%%%%%%%%%%%%%  FIG Two  %%%%%%%%%%%%%%%%%%%%%%%%%%%%
$$\vbox{
\beginpicture
\setcoordinatesystem units <1pt,1pt> point at 0 0
\linethickness 8pt
\arrow <8pt> [.2,.4] from 0 0 to 20 -20
\setlinear \plot 20 -20 40 -40 /
\arrow <8pt> [.2,.4] from 40 -40 to 60 -20
\setlinear \plot 60 -20 80 0 /
\put {$A^L$} [rb] <-1pt,4pt> at 0 0
\put {$\bullet$} at 0 0
\put {$\bullet$} at 80 0
\put {$\bullet$} at 40 -40
\put {$0$} <0pt,-8pt> at 40 -40
\put {$A^R$} [lb] <1pt,4pt> at 80 0
\put {$\gamma_{LR}$} <-4pt,20pt> at 40 -10
\put {$\gamma_{1}$} <-12pt,-10pt> at 20 -20
\put {$\gamma_{2}$} <4pt,-10pt> at 60 -20
\put {{\vbox{\hsize=2.5in \figufont\item{\bf Fig. 2.}
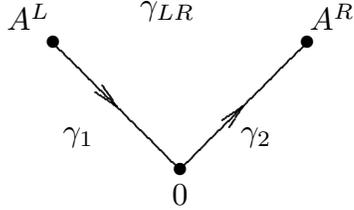
     The path $\gamma_{LR}$ is parametrized by the following
     prescription:
\smallskip
{\halign{\noindent#\hfil\cr
 along $\gamma_1 :~ A(t) = t A^L$~~~~~~~$ t\in [1,0]$\cr
along $\gamma_2 :~ A(t) = s A^R$~~~~~~~$s\in [0,1]$.\cr }}
                          }}} [lb] at 120 -40
\endpicture      }$$
\medskip
%%%%%%%%%%%%%%%%%%%%%%%%%%%%%%%%%%%%%%%%%%%%%%%%%%%%%%%%%%%%%%%%
Then from Eq.\route\ we can obtain $\anom(\Apair ,\gamma_{LR})$ as
\eqn\LRscheme{\eqalign{\anom(\Apair , \gamma_{LR})&=(n+1)\left[
 \int_1^0 dt~ {\rm Tr}\left(
A^L F_L^n(t)\right) +  \int^1_0 ds {\rm Tr}\left(
A^R F_R^n(s)\right)\right]\cr
&\equiv \omega^0_{2n+1,C}(A^R) - \omega^0_{2n+1,C}(A^L)\cr}}
where $\omega^0_{2n+1,C}$ is the Chern-Simon form defined by
\eqn\eq{\eqalign{&\omega^0_{2n+1,C}(A)\equiv (n+1)\int^1_0{\rm Tr}
\left(A\cdot F^n(t)\right)\cr
&{\rm and}~~A(t)=tA~,~~~F(t)=\df A(t) + A(t)^2~.\cr}}
However, in two dimensions we have
\eqn\Chern{\omega^0_{3,C}(A)={\rm Tr}~\left( A\df A\right)+{2\over 3}{\rm Tr}
             \left( A^3\right).}
Thus $\omega^1_{2,LR}$ can be similarly
evaluated from Eq.\crucial\
\eqn\eq{\eqalign{ \df &\omega^1_{2,LR}(\Apair;\vpair)=\var \omega^0_3(\Apair,
       \gamma_{LR})\cr
=\delta_{v_R} \omega^0_{3,C}(&A^R) - \delta_{v_L} \omega^0_{3,C}
              (A^L)
=\df~ {\rm Tr} \left( \df v_R \cdot A^R\right) -
\df~ {\rm Tr} \left( \df v_L \cdot A^L\right)\cr  }}
Therefore,
\eqn\LRanoapp{  \omega^1_{2,LR}(\Apair;\vpair)={\rm Tr}~(\df v_R~ A^R)
-{\rm Tr}~(\df v_L~ A^L)~.}
which is Eq.\LRano .

  {} From Fig.1 and Fig.2, it is  clear that the connection between
the A-scheme and the LR-scheme can be established through a closed-contour
integral.  To show this we first define the integral
\eqn\contour{\df\rho_{2n}(A^{(0)},A^{(1)},A^{(2)})=(n+1)\oint_{\cal C}
     d\tau {\rm Tr}
\left( {\dot A}(\tau)F^n(\tau) \right)   }
where $A(s,t)=A^{(0)} + sA^{(1)} + tA^{(2)}$ and $F(s,t)=\df A(s,t)
    +A^{(2)}(s,t)$.
The contour ${\cal C}$ in the above integral is specified in  Fig.3.
\smallskip
%%%%%%%%%%%%%%%%%%%%%%%%%  FIG Three  %%%%%%%%%%%%%%%%%%%%%%%%%%%%
$$\vbox{
\beginpicture
\setcoordinatesystem units <1pt,1pt> point at 0 0
\linethickness 8pt
\arrow <8pt> [.2,.5] from 80 0 to 40 0
\setlinear \plot 40 0 0 0 /
\put {$\bullet$} at 0 0
\put {$\bullet$} at 80 0
\arrow <8pt> [.2,.5] from 0 0 to 20 -20
\setlinear \plot 20 -20 40 -40 /
\arrow <8pt> [.2,.5] from 40 -40 to 60 -20
\setlinear \plot 60 -20 80 0 /
\put {$\gamma_{1}$} <-12pt,-4pt> at 20 -20
\put {$\gamma_{2}$} <4pt,-10pt> at 60 -20
\put {$\gamma_{3}$} <0pt,10pt> at 40 0
\put {$A^{(0)}$} [rb] <-1pt,-10pt> at 40 -40
\put {$A^{(0)}+A^{(1)}$} [rb] <20pt,4pt> at 95 5
\put {$A^{(0)}+A^{(2)}$} [rb] <10pt,4pt> at 0 5
\put {$\bullet$} at 40 -40
\put {{\vbox{\hsize=170pt \figufont \item{\bf Fig. 3.}
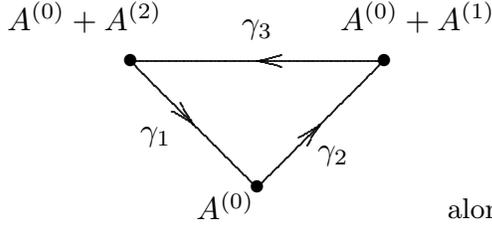
   The contour ${\cal C}$ in Eq.\contour\ is parametrized by the
   following  prescription: \medskip  } }} [lb] at 140 -30
\put {{\vbox{\hsize=180pt \figufont
{\halign{\noindent#\hfil\cr
along $\gamma_1 :  A(t) = A^{(0)} + t A^{(2)}$~~~~~~~$t\in [1,0]$.\cr
along $\gamma_2 : A(s) =A^{(0)}+s A^{(1)}$~~~~~~~$s\in [0,1]$.\cr
along $\gamma_3 : A(u)=A^{(0)}+(1-u)A^{(1)}+uA^{(2)}$~~$u\in [0,1]$.\cr }}
\medskip }}} [lb] at 100 -90
\endpicture        }$$
%%%%%%%%%%%%%%%%%%%%%%%%%%%%%%%%%%%%%%%%%%%%%%%%%%%%%%%%%%%%%%%%
Eq.\contour\ can be evaluated in the following way.
First the integrand can be viewed as an inner-product, in the vector
space spanned by $d{\vec s}$ and $d{\vec r}$.
\eqn\eq{\eqalign{  {\dot A}(\tau)F^n(\tau)&=dsA^{(1)}F^n(s,t) + dt A^{(2)}
F^n(s,t)\cr
&=\left(ds,~dt\right)\cdot \left(A^{(1)}F^n(s,t)~,A^{(2)}F^n(s,t)\right)\equiv
d{\vec s}\cdot {\vec F}~.}}
  {}  From the Stoke's theorem, the closed-contour integral can be
transformed into an area integral
\eqn\eq{\eqalign{ \oint d{\vec s}\cdot {\vec F}&=\int d{\vec a}
\cdot{\vec \nabla}\times {\vec F}\cr
&=\int_{area} da \left( {\partial\over{\partial s}}\left[A^{(2)}F^n(s,t)\right]
-{\partial\over{\partial t}}\left[A^{(1)}F^n(s,t)\right]\right)\cr }}
where the contour specifies the boundary of the area.
This leads to  a straightforward calculation and gives
\eqn\eq{\eqalign{&~~~~~~ \df \rho_{2n}=(n+1)\int^1_0 ds\int^{1-s}_0 dt
\sum^{n-1}_{p=0}\cr
  &~~~~~~~~~ {\rm Tr} \Biggl\lbrace A^{(2)} F^p(s,t)\left[\df A^{(1)}+
A^{(1)}A(s,t)+A(s,t) A^{(1)}\right] F^{n-1-p}(s,t)\cr
  &~~~~~~~~~~~~~~-A^{(1)}F^{n-1-p}(s,t)
\left[\df A^{(2)}+A^{(2)}A(s,t)+A(s,t)A^{(2)}\right]F^p(s,t)\Biggr\rbrace\cr
  &~~~~~~=(n+1)\int^1_0 ds\int^{1-s}_0 dt \sum^{n-1}_{p=0} {\rm Tr} \Biggl
\lbrace A^{(2)} F^p(s,t)\df A^{(1)} F^{n-1-p}(s,t)\cr
  &~~~~~~~~-A^{(1)}F^{n-1-p}\df A^{(2)} F^p(s,t) +A^{(2)}F^p(s,t)A^{(1)}\left[
A(s,t),~ F^{n-1-p}(s,t)\right]\cr
  &~~~~~~~~~~~~~~~~~~~~~~~~~~~~~~~~~~+A^{(2)}
\left[F^p(s,t),~A(s,t)\right]A^{(1)}F^{n-1-p}(s,t) \Biggr\rbrace \cr
  &~~~~~~~=(n+1)\int^1_0 ds\int^{1-s}_0 dt~ \df~~\sum^{n-1}_{p=0}
 {\rm Tr} \left[-A^{(2)}F^p(s,t)A^{(1)}
F^{n-1-p}(s,t)\right]\cr
&\Rightarrow \rho_{2n}(A^{(0)},A^{(1)},A^{(2)})=
(n+1)\sum^{n-1}_{p=0}\int^1_0 ds \int^{1-s}_0 dt~
{\rm Tr}\left(A^{(1)}F^p(s,t)A^{(2)}F^{n-1-p}(s,t)\right)\cr }}
In two dimensions we have
\eqn\Answer{ \rho_2(A^{(0)},A^{(1)},A^{(2)})=2\int^1_0ds\int^{1-s}_0dt~
{\rm Tr} \left(A^{(1)}A^{(2)}\right)={\rm Tr}\left(A^{(1)}A^{(2)}\right)}

Hence the difference between $\anom(\Apair ,\gamma_A)$ and
      $\anom(\Apair ,\gamma_{LR})$ can be obtained if we compare
the contours $\gamma_A$ and $\gamma_{LR}$ and use Eq.\Ascheme\ and \LRscheme\
\eqn\diff{\anom(\Apair , \gamma_{LR})- \anom(\Apair , \gamma_A)=\df\rho_{2n}
(0,A^R,A^L)~.}
Furthermore the differences in the effective action and the anomaly can be
obtained from  Eq.\anomalythree , \crucial\ and \diff\
\eqn\difftwo{\eqalign{
W_{LR} (\Apair) - W_{A} (\Apair)&=c_n \int_{S^{2n}} \rho_{2n}
 (0,A^R,A^L)~,\cr
\omega_{2n,LR}^1(\Apair;\vpair)-\omega_{2n,A}^1(\Apair &;\vpair)
=\var~ \rho_{2n}(0,A^R,A^L).\cr } }

\appendix{B}{}

Following  Ref.\KT , we  explain in some detail how to solve
\eqn\gaueqapptwo{\var \Gamma\left(\Apair,\phi\right)={k\over{8\pi}}
       \int \omega^1_{2,LR} (\Apair;\vpair)~.}
Using the WZ consistency
condition, Eq.\WZcon , Eq.\gaueqapptwo\ can be integrated along
the following path in the
functional space of $\phi, A^L, A^R$:
\eqn\eq{\eqalign{  \phi(x,t)&=g(x,t)\phi(x),~~~~~A^R(x,t)=A^R(x)\cr
A^L(x,t)&=g(x,t)A^L(x)g^{-1}(x,t) - \df_xg(x,t)\cdot g^{-1}(x,t)\cr }}
where $t$ is the parameter specifying the path and $g(x,t)$ satisfies
$g(x,t=0)=${\bf 1} and $g(x,t=1)=\phi(x)^{-1}$.
The particular ingredient in this choice is that one  end
point is $\Gamma\left(\Apair,\phi\right)$, which is what we want, and
the other end point is $\Gamma\left({A^{\prime}}^L,{A^{\prime}}^R,
\phi=1\right)$,
which is easy to evaluate because $\phi=1$.

To be more specific, the infinitesimal variations along $t$
can be evaluated as
\eqn\eq{\eqalign{ \delta \phi(x,t)& = (\Delta t) {dg\over{dt}} g^{-1}
\phi(x,t)\cr
\delta A^L(x,t)&=-(\Delta t) \df_x \left({dg\over{dt}}g^{-1} \right) +
[(\Delta t){dg\over{dt}}g^{-1},~ A^L(x,t)]\cr
\delta A^R(x,t)&= 0~.\cr } }
Together with Eq.\chiralagain\ and \small , we can define the infinitesimal
gauge transformation parameter as $\vraised_L(x)\equiv (\smaDelta t)
{dg\over{dt}}g^{-1}$ and $\vraised_R(x)=0$.
Then the change of $\Gamma\left(A^L(t),A^R(t),\phi(t)\right)$ along the
path can be treated as a result from the gauge transformation specified
by the $\vraised_L(x)$ and $\vraised_R(x)$ above.  Therefore, combining with
Eq.\gaueqapptwo\ we have
\eqn\eq{\eqalign{ \delta &\Gamma\left(A^L(t),A^R(t),\phi(t)\right)=
{k\smaDelta t\over{8\pi}}
\int ~ \omega^1_{2,LR}(A^L(t),A^R(t);{dg\over{dt}}g^{-1},0)~,\cr
\Rightarrow ~~{d\over{dt}}&\Gamma\left(A^L(t),A^R(t),\phi(t)\right)
={k\over{8\pi}}\int
 {}  ~\omega^1_{2,LR}(A^L(t),A^R(t);{dg\over{dt}}g^{-1},0)~.\cr } }
The integration over $dt$ from $0$ to $1$ gives
\eqn\solzero{ \Gamma(A^L,A^R,\phi)=\Gamma(\phi^{-1}A^L\phi+\phi^{-1}
\df\phi,A^R,1)
+{k\over{8\pi}}\int dt \int \omega^1_{2,C}(A^L(t),{dg\over{dt}}g^{-1})}
where we have used Eq.\LRanoapp\ to substitute for $\omega^1_{2,LR}$.
The right-hand side can be solved in the following steps.  We first notice that
if $\phi=1$ and $\vraised_L=\vraised_R=v$,  then $\delta \phi=0$ from
Eq.\small .  Therefore the first term is a solution of
\eqn\eq{ \eqalign{ \delta_{v,v} \Gamma(A^L,A^R,\phi=1)
&={k\over{8\pi}}\int \omega_{2,LR}^1 \left( \Apair; v, v\right)\cr
&={k\over{8\pi}}\int \delta_{v,v} \rho_2(0,A^R,A^L)\cr } }
where we have used Eq.\difftwo\ and the property that $\omega_{2,A}^1(\Apair;
v,v)=0$, and $\rho_2$ is given in \Answer .
Thus we have the solution for the first term on the right hand side,
\eqn\solone{ \Gamma(\phi^{-1}A^L\phi+\phi^{-1}\df\phi, A^R,1)={k\over{8\pi}}
\int {\rm Tr} \left( A^R\phi^{-1}A^L\phi + A^R\phi^{-1}\df\phi \right)~. }
The second term can be simplified by using Zumino's result\Zumino ,
\eqn\soltwo{ \int dt \int_{S^2} \omega^1_{2,C}(A^L(t),{dg\over{dt}}g^{-1})=
{-1\over 3} \int_{B^3} {\rm Tr} \left(\varphi^{-1}\df\varphi \right)^3
+ \int_{S^2} {\rm Tr} \left(-\df\phi \phi^{-1}A^L \right)}
Combining Eq.\solzero , \solone\ and \soltwo\ we obtain
\eqn\eq{ \eqalign{ \Gamma(A^L,A^R,\phi)&={-k\over{24\pi}}
\int_{B^3} {\rm Tr} \left(\varphi^{-1}\df\varphi \right)^3 \cr
  &~~~~~~+{k\over{8\pi}}\int_{S^2} {\rm Tr}
\left( A^R\phi^{-1}\df\phi - \df\phi \phi^{-1}A^L +A^R\phi^{-1}A^L\phi
\right)  \cr }}
This is the solution we quoted in Eq.\gamans .

\appendix{C}{}

In this appendix the evaluation of the determinant
$det({\rm D}^R_z)det({\rm D}^L_{\bar z})$ appearing in Eq.\Jch\ will be shown.
This result was calculated in the A-scheme by many
authors\refs{\PWREF,\Alva,\Schtwo}.  For our purpose it must be appropriately
translated
to the LR-scheme.  In the A-scheme, the result is
\eqn\detR{\left(det{\rm D}^R_z\right)\left(det{\rm D}^L_{\bar z}\right)
\Big\vert_{\rm A-scheme}=\exp\left[2C_2(H)I(h {\widetilde h})\right]\cdot
   \left(det~\delta^{\a\b}\partial_z\right)
   \left(det~\delta^{\a\b}\partial_{\bar z}\right)}
where the $C_2(H)$ is the dual Coxeter number, defined by the structure
constant of the gauge group,
$\sum_{\c \d}f^{\a \c \d}f^{\b \c \d}=-C_2(H)\delta^{\a\b}$.
For $SU(N)$, $C_2(SU(N))=N$. The summation over the indices $\a,\b=
1,2,...,dim~H$.  The vector gauge invariance requires $H_L=H_R=H$.
The appearance of the WZW action indicates the existence of the anomaly
when we transform ${\rm D}_{\mu}$ to $\partial_{\mu}$.
In Sec.3, we need the result in the LR-scheme in which the left gauge
fields are independent of the right ones.  Using Eq.\difftwo ,
we transform the effective action from the A-scheme to the
LR-scheme,
\eqn\eq{\eqalign{ W_{LR}(\Apair)&=W_A(\Apair)+{2C_2(H)\over{4\pi}}\int
  {\rm Tr}\left(A^RA^L\right)\cr
&=2C_2(H)\left[-I(h {\widetilde h})-{1\over{4\pi}}\int d^2z {\rm Tr}\left(
\partial_{z}{\widetilde h}{\widetilde h}^{-1} h^{-1}\partial_{\bar z} h \right)
\right]\cr
&=2C_2(H)\left[-I(h)-I({\widetilde h})\right]\cr  }}
where Eq.\PW\ is used.
Therefore, we have
\eqn\eq{ \eqalign{
\left(det{\rm D}^R_z\right)\left(det{\rm D}^L_{\bar z}\right)
\Big\vert_{\rm LR-scheme}&=\exp\left[2C_2(H) I({\widetilde h})\right]
\left(det~\delta^{\a\b}\partial_z\right)\cr
& ~\cdot~\exp\left[2C_2(H) I({h})\right]
\left(det~\delta^{\a\b}\partial_{\bar z}\right) .\cr}}
Naturally it leads to
\eqn\detRR{\eqalign{ \left(det{\rm D}^R_z\right)\left(det{\rm D}^L_{\bar z}
\right)\Big\vert_{\rm LR-scheme}&=\exp\left[2C_2(H_R) I({\widetilde h})\right]
\left(det~\delta^{\a\b}\partial_z\right)\cr
 & ~\cdot~\exp\left[2C_2(H_L) I({h})\right]
\left(det~\delta^{\a\b}\partial_{\bar z}\right)\cr}}
where $\a ,\b = 1,2,...,dim~H_R$ for the right gauge fields
and $\a ,\b = 1,2,...,dim~H_L$ for the left gauge fields.

\listrefs

\end